\documentclass[letterpaper,fleqn,usenatbib,useAMS]{mnras}

\usepackage{newtxtext,newtxmath}
\usepackage[T1]{fontenc}
\usepackage{ae,aecompl}
\usepackage[pdftex]{graphicx}	
\newcommand{\btg}{G\"{a}nsicke}
\newcommand{\kms}{km\,s$^{-1}$}

\newcommand{\ha}{H$\alpha$}
\newcommand{\hb}{H$\beta$}

\newcommand{\hd}{H$\delta$}

\newcommand{\Porb}{P_{\mbox{\tiny orb}}}
\newcommand{\Msun}{M_\odot}
\newcommand{\Twd}{T_{\mbox{\tiny WD}}}
\newcommand{\Teff}{T_{\mbox{\tiny eff}}}

\newcommand{\be}{\begin{equation}}
\newcommand{\ee}{\end{equation}}

\title[Double-degenerate binaries in spectroscopic surveys]{Using large spectroscopic surveys to test the double degenerate model for Type Ia supernovae}

\author[E. Breedt et al.]{E.~Breedt$^{1}$\thanks{E-mail: E.Breedt@warwick.ac.uk}, 
D.~Steeghs$^{1}$, T.\,R.~Marsh$^{1}$, N.\,P.~Gentile~Fusillo$^{1}$, P.-E.~Tremblay$^{1}$\newauthor
M.~Green$^{1}$, S.~De~Pasquale$^{1}$, J.\,J. Hermes$^{2}$, B.\,T.~\btg$^{1}$, S.\,G.~Parsons$^{3,\,4}$ 
\newauthor M.\,C.\,P.~Bours$^{3}$,  P.~Longa-Pe\~na$^{5}$, A.~Rebassa-Mansergas$^{6}$\\
$^{1}$Department of Physics, University of Warwick, Coventry, CV4 7AL, UK\\
$^{2}$Hubble Fellow, University of North Carolina, Chapel Hill, NC 27599-3255, USA\\
$^{3}$Instituto de F\'isica y Astronom\'ia, Universidad de Valpara\'iso, Avenida Gran Breta\~na 1111,
Valpara\'iso, Chile\\
$^{4}$Department of Physics and Astronomy, University of Sheffield, Sheffield S3 7RH, UK\\
$^{5}$Unidad de Astronom\'ia, Universidad de Antofagasta, Avenida Universidad de Antofagasta 02800, Antofagasta, Chile\\
$^{6}$Departament de F\'isica, Universitat Polit\`ecnica de Catalunya, c/Esteve Terrades 5, E-08860 Castelldefels, Spain\\
}

\date{Accepted XXX. Received YYY; in original form ZZZ}

\pubyear{2017}

\begin{document}

\label{firstpage}

\pagerange{\pageref{firstpage}--\pageref{lastpage}}

\maketitle

\begin{abstract}
An observational constraint on the contribution of double degenerates to Type~Ia \nobreak{supernovae} requires multiple radial velocity measurements of ideally thousands of white dwarfs. This is because only a small fraction of the double degenerate population is massive enough, with orbital periods short enough, to be considered viable Type~Ia progenitors. %
We show how the radial velocity information available from public surveys such as the Sloan Digital Sky Survey can be used to pre-select targets for variability, leading to a ten-fold reduction in observing time required compared to an unranked or random survey. %
We carry out Monte Carlo simulations to quantify the detection probability of various types of binaries in the survey and show that this method, even in the most pessimistic case, doubles the survey size of the largest survey to date (the SPY survey) in less than 15 per cent of the required observing time. %
Our initial follow-up observations corroborate the method, yielding 15 binaries so far (eight known and seven new), as well as  orbital periods for four of the new binaries.

\end{abstract}



%
\section{Introduction} \label{sec:intro}

Type Ia supernovae (SNe\,Ia) play a central role in modern astrophysics. They are among the brightest explosions in the universe, visible across extragalactic distances, and they are responsible for the synthesis of heavy elements \citep{wiersma11} and the acceleration of cosmic rays \citep{schure13}. 
Their light curves display a tight correlation between the peak luminosity and brightness evolution \citep{phillips93}, which make them useful as `calibrated candles' to track the expansion history of the universe. The discovery of the accelerated expansion of the universe \citep{riess98, perlmutter99}, for which the 2011 Nobel Prize in physics was awarded, is fundamentally based on observations of large numbers of SNe\,Ia.

It is generally accepted that a SN\,Ia explosion occurs as a result of runaway thermonuclear fusion in a massive~($\gtrsim~1\Msun$) carbon-oxygen white dwarf, but the exact mechanism of the explosion, the nature of the progenitor binary and the evolutionary pathways that lead to such a massive white dwarf are less clear (see \citealt{maoz14} and \citealt{postnovyungelson14} for detailed reviews). In broad terms, two classes of progenitor models exist. In the `single-degenerate' case \citep{whelaniben73}, the white dwarf accretes hydrogen-rich material from a non-degenerate companion until compressional heating ignites the carbon and triggers the explosion. In the `double-degenerate' model \citep{webbink84, ibentutukov84} both stars are white dwarfs. They spiral together due to gravitational wave emission and the associated angular momentum loss until they eventually merge. Those systems with sufficient combined mass and orbital periods shorter than $P\la12$\,h, implying a merger time less than a Hubble time, are assumed to be SN\,Ia progenitors. Therefore, even though double white dwarf binaries are abundant \citep[e.g.][]{nelemans05,napiwotzki07,holberg16}, only a small fraction of the total population are potential SN\,Ia progenitors. 
When the explosion mechanism and evolutionary timescales are considered as well, there are a number of variations on these two progenitor models. For example, binaries containing a massive white dwarf plus a hot subdwarf star (WD+sdO/sdB) are recognised as potential double degenerate SN\,Ia progenitors as well, since subdwarf lifetimes are short or comparable to the gravitational merger timescale \citep[e.g.][]{michaud11,heber16}. The WD+sdB binary KPD\,1930+2752 \citep{maxted00KPD,geier07} with its 136.9\,min orbital period and combined mass of between $1.36-1.48\Msun$, is considered as one of the best candidates for a future double degenerate SN\,Ia explosion. Other model variations include stellar rotation or magnetic fields as stabilising factors, resulting in super-Chandrasekar supernovae \citep[e.g.][]{scalzo12}. Theoretical models and hydrodynamical simulations also suggest that ``double detonations'' can occur, causing sub-Chandrasekhar mass white dwarfs to explode as SNe\,Ia. This happens because of off-centre explosive ignition of the accreted helium layer, which sends a shock wave through the star and detonates the carbon-oxygen core \citep[e.g.][]{woosleyweaver94, fink10, sim10}. This wide range of models and simulation results imply that a variety of explosion masses and progenitor types are possible, and that several progenitor types may contribute to the observed SN\,Ia rate. What their relative contribution is or whether one channel dominates the production of SN\,Ia, is still unknown. In the future it will likely be these systematic differences, rather than the size of the statistical SN\,Ia sample or calibration uncertainties, that will limit the accuracy of SN\,Ia cosmology.

At the moment, both binary population synthesis calculations and observations slightly favour the double degenerate model. The best-observed supernovae also show no signs of a surviving companion, suggesting that a merger took place \citep[e.g.][]{schaeferpagnotta12}. But there are also observations that point to a single degenerate origin for the explosion. For example, \citet{cao15} interpret an ultraviolet flash seen four days after the explosion as the interaction of the material ejected from the supernova with the surviving companion star. Binary population synthesis models find that the observed delay time distribution (i.e. the distribution of times for binaries to merge and explode as SNe\,Ia) matches the expected distribution for double degenerates and gravitational wave radiation more closely than can be produced with single degenerate models  \citep{mennekens10,maoz14,yungelson17}. These population models produce a white dwarf merger rate that is close to the observed Milky Way SN\,Ia rate, $5.4\times10^{-3}$ per year \citep{li11}, but they are limited by the small number of well-studied double degenerate binaries available to calibrate these models with, and in particular, the biases in the observed samples \citep{toonen12}. The currently known sample of double degenerates is heavily skewed towards the extremely low mass white dwarfs \citep[ELM;][]{brown16elm7}. These are pairs of helium white dwarfs where at least one component has $M\lesssim0.3\Msun$. As the Universe is not old enough for them to have formed via single star evolution, they have to be the product of binary common envelope evolution. The high mass end, where the potential SN\,Ia progenitors lie, is still largely unconstrained, so the contribution of the various types of progenitor binaries to the SN\,Ia rate is still unknown. 

One of the key criteria a potential progenitor population has to meet is that there has to be enough of them to account for the observed SN\,Ia rate. There are large uncertainties in the parameters contributing to this figure, such as the total stellar mass in the Galaxy, its star formation history and the white dwarf binary fraction. Population synthesis estimates of the median white dwarf merger timescale in the Galaxy is $\sim 0.7 - 1$\,Gyr  \citep{ruiter11,toonen12,yungelson17}. Using the observed Milky Way SN\,Ia rate, roughly $4.6\times10^6$ SNe\,Ia will be observed over this time. With $\sim10^9$ white dwarfs in the Galaxy \citep{harris06}, we should expect to find at most one in $\sim200$ observed white dwarfs to be a SN\,Ia progenitor, if SN\,Ia are produced solely by double degenerate binaries. Allowing for a factor of $\sim2$ systematic uncertainty in the observed SN\,Ia rate \citep{li11} and less than perfect detection efficiency of follow-up observations, the number could easily be half this estimate, or less. On the other hand, using the local stellar mass to white dwarf ratio, \citet{maoz11} estimate that as many as 1 in 40 white dwarfs could be SN\,Ia progenitors. Despite these large uncertainties, it is clear that for a robust observational estimate of the double degenerate contribution to SN\,Ia we require a survey of at least a few thousand white dwarfs. 

The lack of direct interaction or accretion make detached white dwarf binaries difficult to find. Their spectra generally resemble single white dwarfs, either because the light from one of the white dwarfs dominates the spectrum, or because the spectra of two hydrogen white dwarfs cannot easily be separated at low to moderate spectral resolution.
In most cases, the presence of the companion is only inferred from radial velocity variations of the more luminous white dwarf. The radial velocity amplitude of a typical SN\,Ia progenitor binary is $>100$\,\kms, so there has been a number of attempts at identifying new double degenerates by  surveying known white dwarfs for radial velocity variability. Several double degenerates have been found this way (see \citealt{napiwotzki04} for a detailed summary of previous surveys and their results), but none massive enough to be a SN\,Ia progenitor. Individual examples of sufficiently massive double degenerate binaries have been discovered \citep{santandergarcia15}, but to be able to assess the supernova rate that could potentially come from double degenerates, ideally thousands of white dwarfs need to be surveyed.

The most recent and by far the largest of these dedicated surveys was the SN\,Ia Progenitor SurveY \citep[SPY,][]{napiwotzki07}. SPY targeted  1014 catalogued white dwarfs brighter than $B,V<16.5$, taking multiple radial velocity snapshots using the high resolution spectrograph UVES on the European Southern Observatory's Very Large Telescope \citep{napiwotzki01,napiwotzki07}. The survey discovered 39 double degenerate binaries, along with many subdwarf-B stars \citep[sdB;][]{geier08}, subdwarf-O stars \citep[sdO;][]{stroeer07} and white dwarfs with low mass companions \citep[e.g.][]{maxted07}. The survey included a total of 615 hydrogen (DA) white dwarfs which were tested for radial velocity variability \citep{koester09}. Further follow-up observations are needed to measure the parameters of the newly identified binaries, but at least one $\Porb=9.06$\,h sdB+WD binary was found to have a total mass close to the Chandrasekhar limit \citep{geier10}, and one double white dwarf system has a best-estimate total mass of $1.35\Msun$ \citep{napiwotzki07,geier07,maoz14}. Although these two detections in a survey of $\sim$1000 targets are broadly consistent with the expected number of progenitors for the double degenerate SN\,Ia channel, it is clear that larger samples of white dwarfs are needed to yield a conclusive result. 

SPY produced valuable results for a range of white dwarf-related science (see e.g. \citealt{geier11spy} and references therein), but a dedicated survey like this is observationally expensive. Our goal now is to investigate whether using the data from public spectroscopic surveys is a viable alternative, i.e. whether a large number of double degenerates can be identified in this way. The Sloan Digital Sky Survey \citep{sdsstech_york00} includes spectra of tens of thousands of white dwarfs in its public database. Although much less precise than SPY, the individual exposures that the spectra are composed of (see Section~\ref{sec:subspec}) contain useful radial velocity information. Several groups have used this information to identify or characterise white dwarf binaries, e.g. white dwarf-main sequence binaries \citep{rebassa07,nebot11,parsons13}, AM\,CVn stars \citep{carter14}, WD+sdB binaries \citep[the MUCHFUSS project;][]{geier11,kupfer15} and double white dwarf binaries \citep{mullally09, badenes09}. At least one relatively massive double white dwarf binary has been discovered this way so far \citep{kulkarni10,marsh11sdss1257}. Additionally, \citet{badenes12} used the SDSS radial velocity information to calculate the merger rate of white dwarf binaries in the Galaxy. They measure the maximum radial velocity change observed for each white dwarf in their sample and compare the distribution of velocities with that of various simulated white dwarf binary populations. They show that the general white dwarf merger rate is comparable with the SN\,Ia rate in the Galaxy, but only if massive, sub-Chandrasekhar white dwarfs contribute to the SN\,Ia rate as well. If only Chandrasekhar mass mergers are considered, the rate is an order of magnitude too low. \citet{yungelson17} reach a similar conclusion based on detailed binary population synthesis modelling. 

In this paper we use the SDSS velocities to create a priority ranking system for follow-up of binary candidates. To demonstrate the method, we restrict our analysis to hydrogen atmosphere (DA) white dwarf binaries only, as these are by far the most common subtype of white dwarf \citep[see e.g.][]{girven11,gentilefusillo15} and also have the simplest spectra. An extension of the survey to include helium atmosphere white dwarfs and sdB stars would be straightforward. We use Monte Carlo simulations to quantify the detection efficiency of the SDSS spectra for binaries with range of parameters (Section ~\ref{sec:sims}) and then show the gain in survey efficiency of such a ranked survey compared to blind targeting of the full sample. We present some exploratory observations in Section~\ref{sec:obs} and discuss the application of the method to future large scale spectroscopic surveys in Section~\ref{sec:discussion}.

%
\section{White dwarfs in SDSS} \label{sec:subspec}

\begin{figure}
 \centerline{\includegraphics[width=0.80\columnwidth]{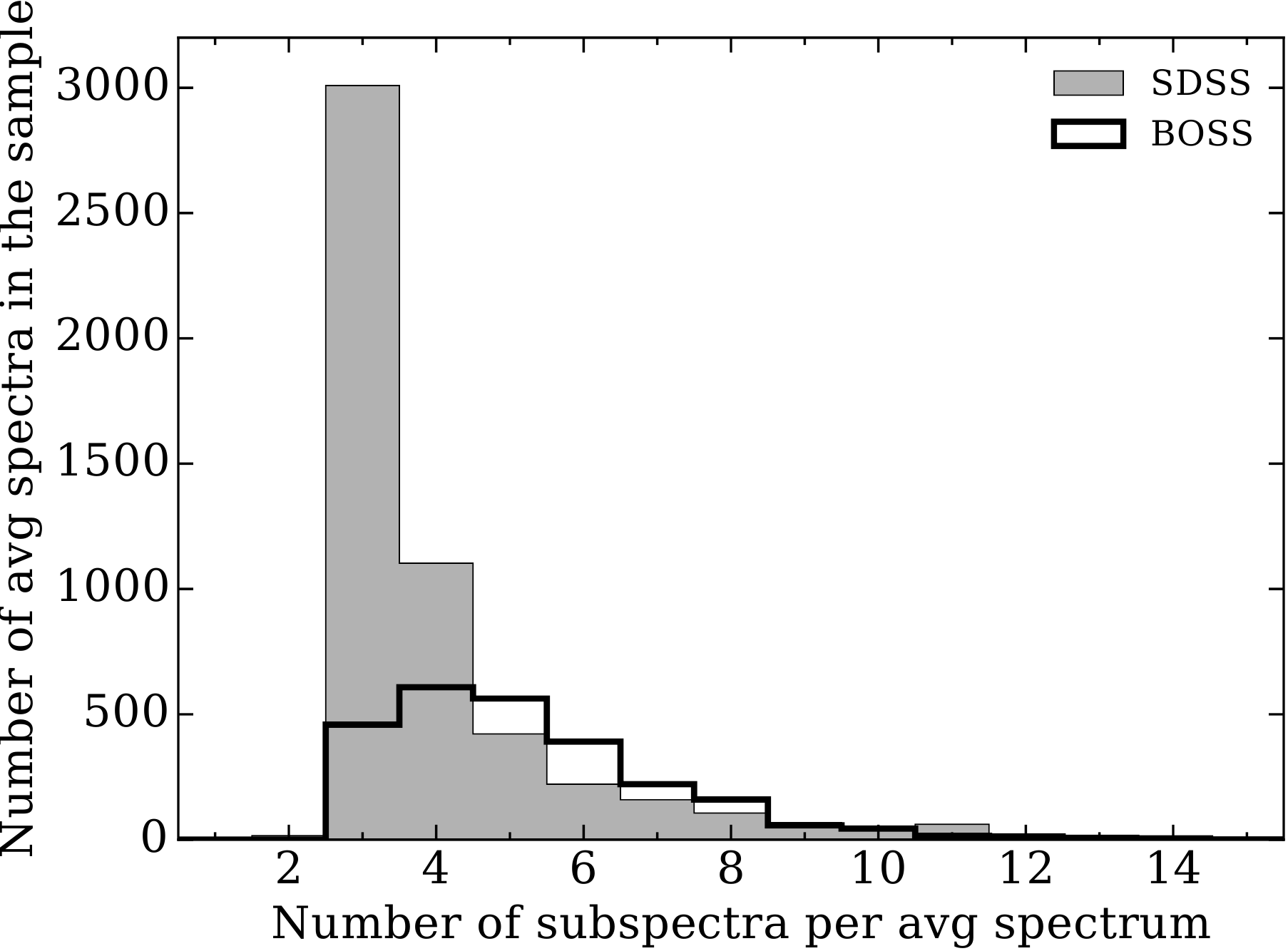}}
 \vspace{1mm}
 \centerline{\includegraphics[width=0.82\columnwidth]{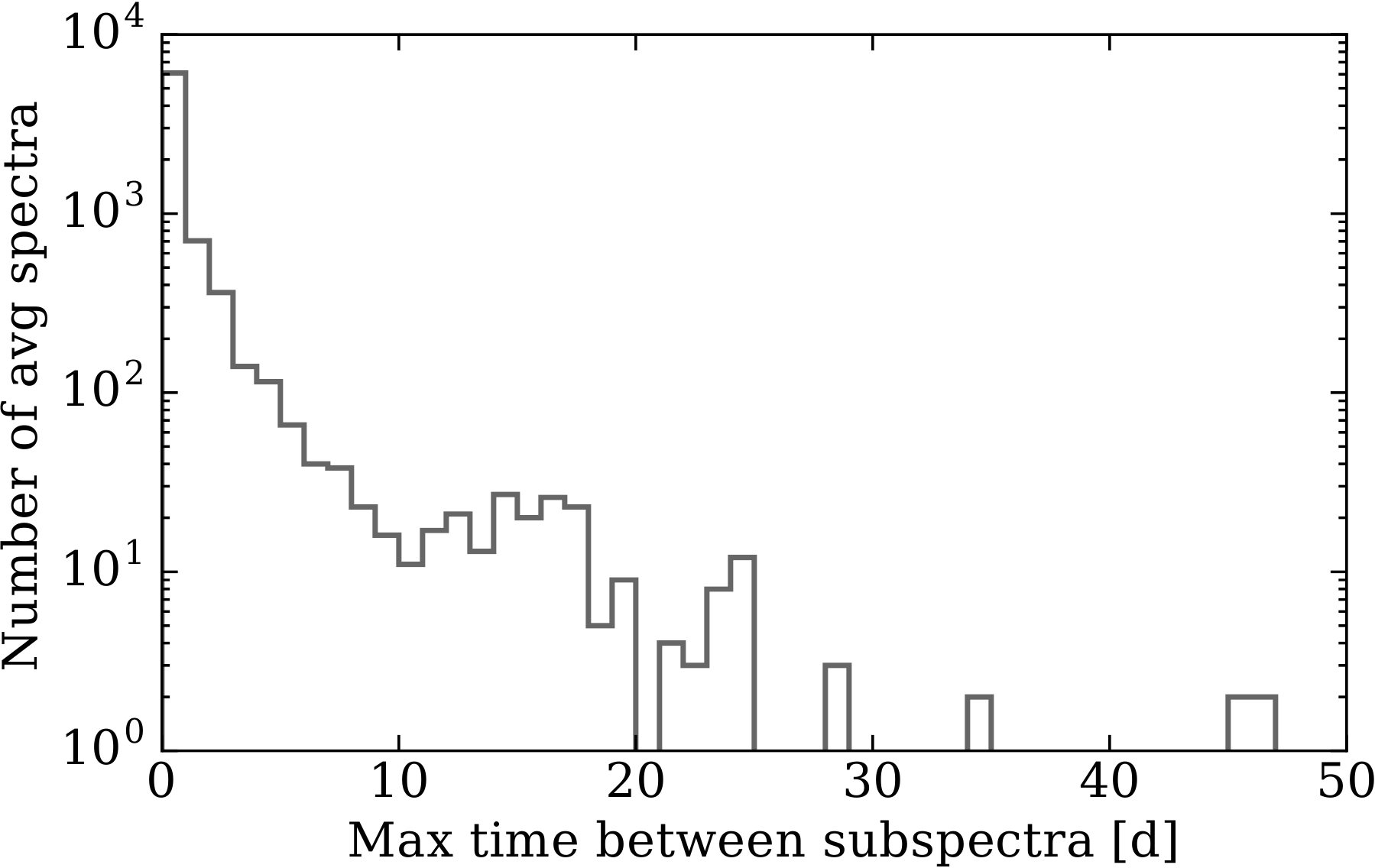}}
 \caption{{\em Top:} Number of subspectra for each average spectrum in our sample. The SDSS and BOSS surveys are shown separately. 1122/6396 of our targets have spectra from both surveys. Each of these subspectra provide a radial velocity snapshot of the target. {\em Bottom:} Histogram of the time between the earliest and latest subspectrum contributing to a single average spectrum. For the vast majority of targets the observations are completed within a single night, but sometimes additional exposures are needed to reach the desired signal to noise ratio in the average spectrum.}
 \label{fig:subspectra}
\end{figure}

The Sloan Digital Sky Survey (SDSS) has accumulated tens of thousands of spectra over the last decade as part of its various science projects. The original survey instrument was a fibre-fed spectrograph with 640 fibres per plate, which produced spectra covering $3800-9200$\,\AA\, at a resolution of $\sim1500-2500$ over the wavelength range \citep{sdsstech_york00}. The spectrograph was rebuilt and upgraded at the end of 2008, and is now known as the Baryon Oscillation Spectroscopic Survey (BOSS) spectrograph. It has 1000 fibres, and its blue and red channels cover $3600-6350$\,\AA\, and $5650-10\,000$\,\AA\, respectively, at a similar resolution to the original SDSS spectrograph \citep{sdssdr9}.

We started from a sample of 7958 spectra taken from SDSS Data Release 10, which were classified by \citet{gentilefusillo15} as non-magnetic DA  white dwarfs. This number includes duplicates, as target selection strategies in the SDSS and BOSS survey programmes are independent of each other, and a given target may have multiple spectra if it was selected for spectroscopy by more than one programme. 1122 of our targets have spectra taken with both the SDSS and the BOSS spectrographs. The white dwarf selection is based on a wide region of colour space, defined to include all spectroscopically confirmed and candidate white dwarfs in SDSS. All SDSS spectra within this colour space were then visually inspected and classified, and from this we selected only the confirmed DA white dwarfs for our double degenerate search. A magnitude limit of $g\leq19.0$ was applied to the full selection to be able to ensure a reliable classification. We find that this limit is also appropriate for the radial velocity work we do here, as beyond this limit the signal-to-noise ratio (SNR) of the subspectra is generally too low for a reliable radial velocity measurement. A typical average spectrum at this magnitude has a SNR of $\sim10$. Because of the visual inspection step we believe the sample to be complete for all $g\leq19.0$ DA white dwarfs in Data Release 10 which have SDSS spectroscopy. Note, however, that SDSS is only $\sim40$ per cent complete in its spectroscopic selection of white dwarfs \citep{gentilefusillo15}. 

Since we are interested specifically in the double degenerate population, we first cleaned the sample from known binaries of other types. In particular, some white dwarf--main sequence (WDMS) binaries are entirely dominated by the white dwarf at optical wavelengths, so look like single white dwarfs in SDSS. Their late type companions are revealed at redder wavelengths. We removed all known WDMS binaries from our sample by cross-matching against the latest SDSS WDMS catalogue \citep{rebassa12,rebassa16}. We also removed white dwarfs known to have a photometric infrared excess \citep{girven11}, which could indicate the presence of a late-type companion or a dust disk, as well as any other white dwarfs for which weak emission lines were detected in the average spectra. Finally, we also removed spectra which had problems (such as a missing part of the spectrum) if this was severe enough to affect the radial velocity measurements. The final, cleaned sample contains 7792 spectra, corresponding to 6396 unique white dwarfs.

Each SDSS spectrum is the average of three or more \nobreak{{\em subspectra}}: individual 900\,s exposures which are repeated until the signal-to-noise of the combined spectrum reaches the SDSS survey requirement\footnote{http://classic.sdss.org/dr7/products/spectra/}. At our $g\leq19.0$ cut-off the subspecta have a typical SNR of $\sim3-4$. The number of subspectra contributing to an average spectrum in our white dwarf sample, as well as the time span of the observations, are shown in Figure~\ref{fig:subspectra}. For most targets, the required signal-to-noise is achieved with three consecutive exposures on a single night, but sometimes further observations are needed, and these may be taken on different nights. The 6396 white dwarfs in our sample have a total of 34\,446 subspectra from which radial velocity information can be obtained.

\subsection{White dwarf masses} \label{sec:wdmass}

For each of the 6396 targets we first fit the average spectrum with a grid of atmosphere models covering $\Teff=1500-140\,000$~K in temperature and $\log g=6.5-9.5$, where $g = GM/R^2$ is the surface gravity of the white dwarf. The calculation of the model grid as well as the fitting procedure are described in detail in \citet{tremblay11}. We also include the correction required for 3D effects at low temperatures \citep{tremblay11_3d, tremblay13}. For targets with more than one average spectrum, we adopted the parameters derived from the spectrum with the highest signal to noise ratio. To convert the best-fit $\log g$ to white dwarf mass we use the mass-radius relations of \citet{fontaine01} for white dwarfs with $\Teff<30\,000$~K and \citet{wood95} if $\Teff>30\,000$~K.

The spectra of most double degenerates are indistinguishable from single white dwarfs. The spectra are often dominated by the more luminous white dwarf, and even if both white dwarfs contribute significantly, the spectra still resemble single white dwarfs, especially at the spectral resolution of SDSS. In the standard picture of stellar evolution, the more massive star in a binary will become the more massive white dwarf, and it will reach this phase of evolution sooner than its companion. So, as a double degenerate binary, the more massive white dwarf is typically the fainter of the pair, both because of its smaller radius and because it has been cooling for longer. The luminosity is therefore dominated by the youngest, least massive white dwarf of the pair, so it is this star that the parameters derived from the model atmosphere fits refer to. It is possible for the more massive white dwarf to be the more luminous of the pair, but such cases are much rarer. It requires the binary to have evolved through a period of conservative mass transfer during which the initially more massive (and hence more evolved) star lost most of its mass in a common envelope event and becomes the least massive white dwarf. When the companion star becomes a white dwarf it is the more massive of the pair and because it is hotter, it dominates the luminosity \citep{moran97,rebassa17}. \citet{toonen12} show that although different models of the common envelope evolution produce different distributions of the mass ratio $q$, the vast majority have $q=M_1/M_2 < 1$, clustering around $q\sim0.5$. So, for simplicity, we do not include these rare, inverted mass ratio binaries in our simulations. 

\citet{tremblay11} show that DA white dwarfs with helium white dwarf companions (DC or DB) can usually be recognised as binary candidates from the discrepancy between their temperatures derived from spectroscopy and from multi-band photometry. DA+DB or DA+DC binaries can also be recognised from their combined spectra which display both hydrogen and helium lines (spectral type DAB or DBA) or from unusual hydrogen line profiles. For DA+DA binaries this distinction is less clear, so we have to rely on detecting radial velocity variations.

\subsection{Radial velocities}

\begin{figure}
 \centerline{\rotatebox{270}{\includegraphics[height=0.8\columnwidth]{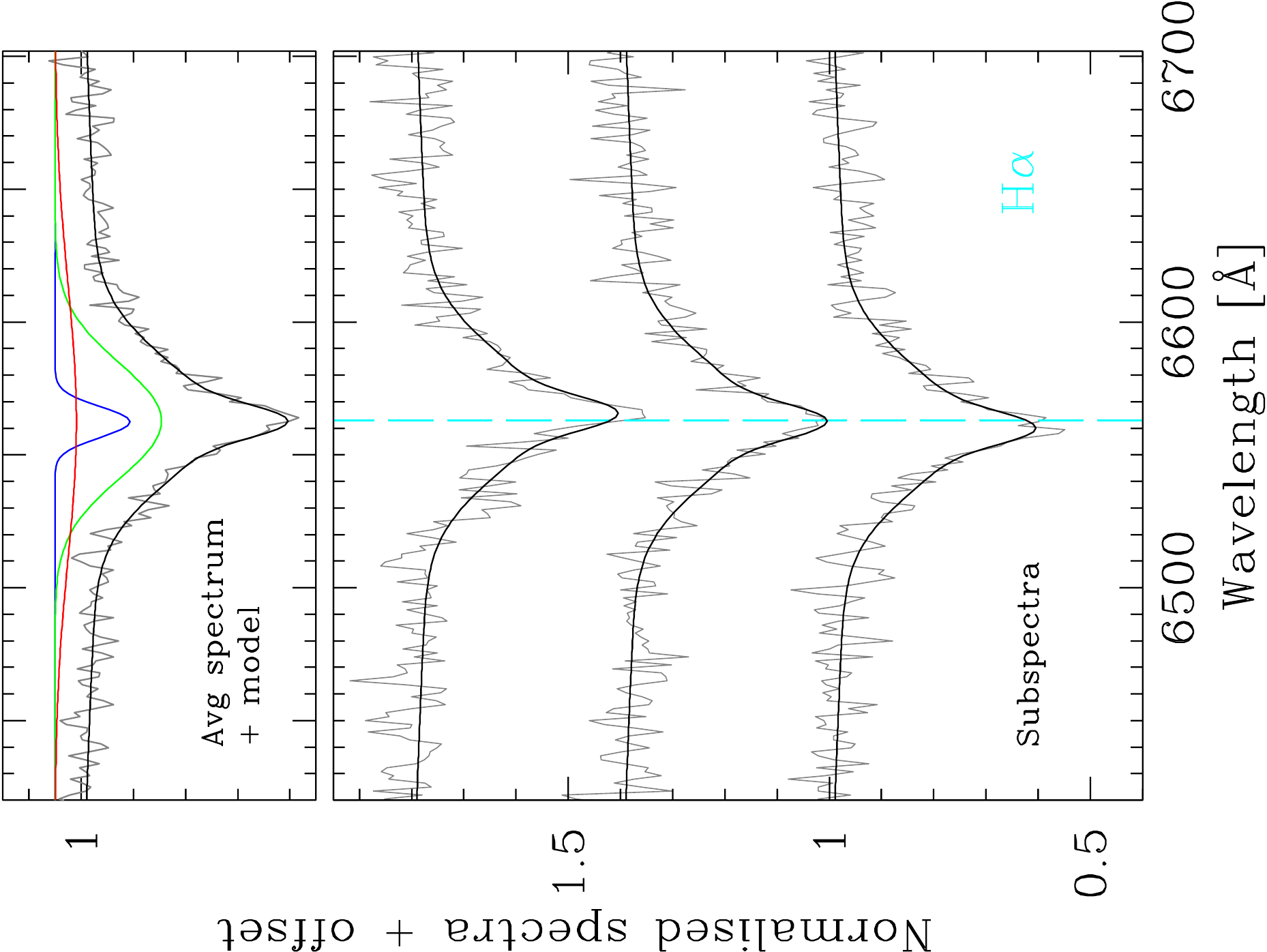}}}
 \caption{Triple Gaussian fit to the subspectra of SDSS\,J123549.86+154319.0. A shift from blue to red around the rest wavelength of \ha\, (vertical dashed line) is clearly visible. The widths and depths of the Gaussian functions are derived from a fit to the average spectrum (top panel). For the subspectra, those parameters are fixed and only the position along the wavelength axis is allowed to vary. The \ha\, to \hd\, lines are fit simultaneously and are constrained to have the same radial velocity shift. }
 \label{fig:subspecfit}
\end{figure}

We measured the radial velocities in two different ways. First we carried out a multi-Gaussian fit to each of the four Balmer lines, \ha\, to \hd, in the continuum-normalised \nobreak{{\em average}} spectra. 
Each of the Balmer lines were modelled as the sum of three or two Gaussian functions, for a signal-to-noise above or below 20, respectively. This allowed us to model both the broad wings and the narrow core of the lines accurately. For each line, we restricted the width of the narrowest Gaussian in the model to be wider than the spectral resolution. An example fit is shown in Figure~\ref{fig:subspecfit}. The four lines (\ha\, to \hd) were fit simultaneously, constrained to all have the same radial velocity shift. We then fixed all the parameters at their best-fit values, leaving only the radial velocity free, and fit this model to each of that target's subspectra. This has the advantage that the low signal-to-noise subspectra can be reliably measured and that noise spikes are not over-fitted by a model with many free parameters. Cosmic rays were removed during the fit using iterative sigma clipping. The fitting routine is a non-linear least squares minimization, implemented using the Python package {\sc lmfit} \citep{lmfit} and the quoted errors are standard $1\,\sigma$ errors on the fitted parameters. 

As another check on our velocities, we used the best-fit white dwarf atmosphere model for each target (section~\ref{sec:wdmass}) as a template and fit it to each of the subspectra. We calculated the $\chi^2$ of the fit for a range of small steps in radial velocity and then recorded the velocity at the minimum of the resulting $\chi^2$ curve. The error on the radial velocity can be estimated using bootstrapping techniques or from the width of the $\chi^2$ curve around the minimum \citep[e.g.][p.\,692]{numericalrecipes}. The two fitting methods yield entirely consistent results, but we found that especially in the case of hot, massive white dwarfs with broad, shallow absorption lines, the $\chi^2$ changed slowly when using the second method, leading to very broad $\chi^2$ curves and hence large errors on the radial velocity. In general we prefer the Gaussian fitting method for the ease and reliability with which the error on the velocity measurement can be determined. 

The SDSS spectra are already corrected to the solar system barycentre, so we apply no further corrections to the times or the measured velocities.

%
\section{Double degenerate candidate selection} \label{sec:method}

We quantify the variability of our targets using a variability parameter $\eta$ as described by \citet{maxted00}. It is calculated as follows. For each target we calculate the error-weighted mean of its radial velocities, using all $n$ available subspectra for that target. We then calculate a $\chi^2$ value, $\chi^2_m$, against the assumption of constant radial velocity at this mean value.  %
Comparing the observed $\chi^2_m$ against a $\chi^2$ distribution with $n-1$ degrees of freedom, gives the probability $p$ of measuring an equal or larger $\chi^2$ value from $n$ random variates. So, if the set of measured radial velocities is very different from a distribution that can be described as random scatter around a mean value (e.g. large differences between the individual radial velocity measurements or a set of steadily increasing velocity values) $\chi^2_m$ is large and it will fall in the tail of the $\chi^2$ distribution. Only a very small fraction of random fluctuation outcomes can produce a $\chi^2$ as large as or larger than $\chi^2_m$, so the probability that the measured variation is due to noise only, is low. For convenience, we use the logarithm of this probability and define the variability parameter %
\be \eta = -\log_{10} [P(\chi^2 > \chi^2_m)], \label{eq:eta} \ee %
so the larger the value of $\eta$, the smaller the probability that the measured radial velocity variability is due to chance.

\begin{table*} 
  \centering
  \caption{\label{tab:known} Known binaries in the sample, with measured orbital parameters, arranged by the variability parameter $\eta$}
  \begin{tabular}{cccccc} \hline
     Target                 & $\eta$ (SDSS) & $\Porb$ (h) & $M_1/M_\odot$ & $M_2/M_\odot$ &  Ref. \\
  \hline\hline
     SDSS\,J075519.47$+$480034.1  &  100.0  & 13.11048      & 0.410     & $>0.89$    & $a$ \\
     SDSS\,J092345.59$+$302805.0  &  100.0  & 1.0788        & 0.274     & $>0.37$    & $b$\\
     SDSS\,J125733.64$+$542850.5  &  50.47  & 4.55500(2)    & 0.15(5)   & $0.92(13)$ & $c,d,e$\\
     SDSS\,J143633.29$+$501026.7  &  23.16  & 1.0992        & 0.233     & $>0.45$    & $f,g$\\
     SDSS\,J082511.90$+$115236.4  &  21.97  & 1.39656       & 0.278     & $>0.49$    & $h$\\
     SDSS\,J100559.10$+$224932.3  &   5.23  & 2.7843704(36) & 0.378(23) & 0.316(11)  & $i,j,k$\\
     SDSS\,J112721.28$-$020837.4  &   4.15  &     ~         & 0.565(7)  &  ~         & $l$\\       
     SDSS\,J105353.89$+$520031.0  &   3.56  & 1.02144       & 0.204     & $>0.26$    & $f,b$\\
     SDSS\,J081544.25$+$230904.7  &   1.13  & 25.76568      & 0.200     & $>0.50$    & $a$\\
     SDSS\,J155708.48$+$282336.0  &   0.30  & 9.77784       & 0.494     & $>0.43$    & $a$\\
     SDSS\,J123410.36$-$022802.8  &   0.27  & 2.19432       & 0.227     & $>0.09$    & $g$\\
     SDSS\,J143042.61$+$371015.0  &   0.16  & 27.76176(5)   & 0.348     & $>0.233$   & $m$\\
     SDSS\,J225242.25$-$005626.6  &   0.12  &     ~         & 0.349     &    ~       & $g$\\
     SDSS\,J110436.70$+$091822.1  &   0.10  & 13.27656      & 0.457     & $>0.55$    & $a$\\
  \hline\\
  \end{tabular}
~\\{\em References:} $a -$\citet{brown13elm5}, $b -$\citet{brown10elm1}, $c -$\citet{badenes09}, $d -$\citet{kulkarni10}, $e -$\citet{marsh11sdss1257}, $f -$\citet{mullally09}, $g -$\citet{kilic11elm2}, $h -$\citet{kilic12elm4}, $i -$\citet{parsons11}, $j -$\citet{bours14}, 
$k -$\citet{bours15}, $l -$\citet{maoz17}, $m -$\citet{moralesrueda05}
\end{table*}

Table~\ref{tab:known} shows the parameters of the 14 known double degenerate binaries in our sample, arranged by the value of $\eta$ we measure from their SDSS subspectra. Only half of these binaries have $\eta$ high enough to clearly identify them as binaries or binary candidates ($\eta>2.5$; see Section~\ref{sec:discussion}). In practice, the value of $\eta$ we measure from the sparsely sampled radial velocity observations, depends on the orbital period of the binary and the time between exposures. The typical SDSS exposure pattern (three subsequent spectra taken over $\sim45$~minutes) is less sensitive to binaries with long orbital periods or low mass companions, as such binaries will display only a small change in radial velocity over the duration of the observation. But since the exposure pattern is not fixed, and many targets have additional exposures on another day (Figure~\ref{fig:subspectra}), long-period binaries are not completely excluded from the sample. In the next section we use Monte Carlo simulations to quantify the detection efficiency for various types of binaries in SDSS, using a range of component masses and orbital periods.

\begin{figure}
 \centerline{{\includegraphics[width=\columnwidth]{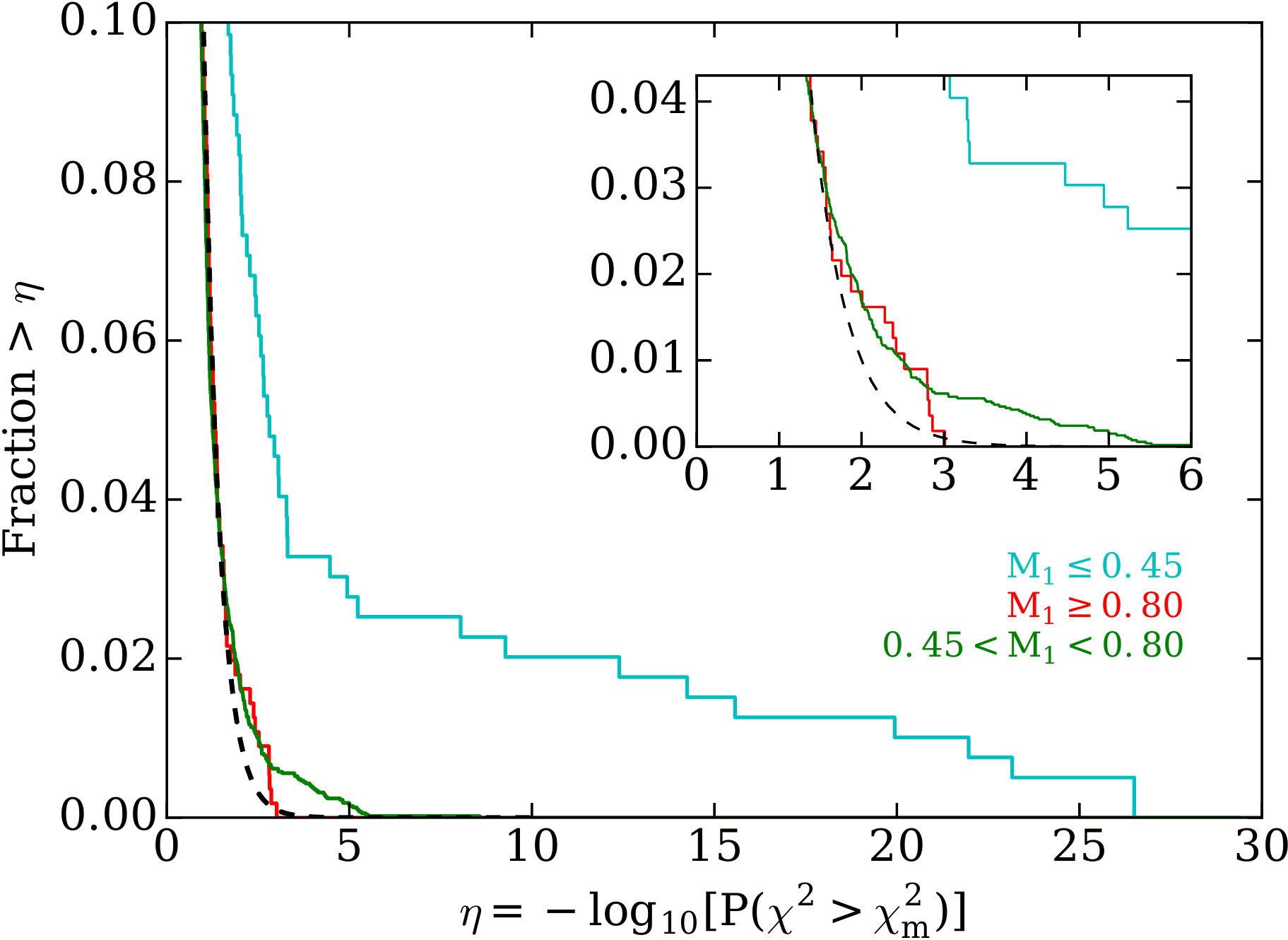}}}
 \caption{Radial velocity variability of the SDSS white dwarf sample, separated by mass as shown. The variability parameter $\eta$ on the horizontal axis is as defined in Section~\ref{sec:method}, and the vertical axis represents the fraction of the sample that displays variability $\mbox{$>\eta$}$. The black dashed line is the prediction for a sample of pure noise. All three groups contain targets which display variability in excess to random noise. The small panel shows an expanded version of the plot for clarity.}
 \label{fig:psplitmass}
\end{figure}

First, we consider the variability of the SDSS white dwarf sample as a whole. Sorted by $\eta$, 11 of the top 20 targets are known or newly confirmed binaries (Section~\ref{sec:obs}). The other nine are still awaiting follow-up observations, but at least seven display clear, likely long-period, radial velocity variability in their subspectra. Figure~\ref{fig:psplitmass} shows the cumulative distribution function of the variability parameter $\eta$, separated into three groups by mass. There are 556 massive white dwarfs ($M\geq0.8\Msun$; red line) in the sample, 5363 white dwarfs with masses $0.45\Msun< M < 0.8\Msun$ (green line) and 396 low mass white dwarfs ($M\leq0.45\Msun$, cyan line). Only three white dwarfs in the low mass group have $M<0.20\Msun$, so this group is perhaps better described as $0.20\Msun\lesssim M < 0.45\Msun$. This is likely a consequence of the colour region where our sample was selected from (Section~\ref{sec:subspec}). We excluded 81 targets for which the mass could not be reliably measured from the SDSS spectra. 
The dashed line is $10^{-\eta}$, corresponding to a sample of pure noise. At low values of $\eta$, the variability of the $M\geq0.80\Msun$ and $0.45\Msun< M < 0.80\Msun$ groups match the prediction for random noise, but above $\eta\gtrsim1.5$ both groups show an excess. 

The low mass white dwarfs join the noise curve for $\eta\lesssim0.4$ and show a pronounced excess above the noise for values higher than that. Low mass white dwarfs are known to have a high binary fraction \citep{bergeron92,marsh95,brownj11,rebassa11,brown16elm7}, since single stars cannot evolve to such low mass white dwarfs within a Hubble time. Instead these white dwarfs are formed as a result of enhanced mass loss to a binary companion, so a large variability excess above the noise curve should be expected for this mass range. For example, \citet{brownj11} find a $70$~per~cent binary fraction for white dwarfs with $M\leq0.45\Msun$, increasing to $>95$~per~cent for $M<0.20\Msun$. The ELM survey find that the masses of the companion white dwarfs follow a normal distribution with $M_2 = (0.76\pm0.25)\Msun$, leading to an average total mass $M = (1.01\pm0.15)\Msun$ \citep{brown16elm7}, so it is unlikely that we will find many SN\,Ia progenitors in this group. These binaries are much more likely to merge and become more massive single white dwarfs, R~Coronae~Borealis stars or mass-transferring AM\,CVn binaries \citep{webbink84,marsh04}, but the excess in Figure~\ref{fig:psplitmass} is a clear illustration of the sensitivity of the SDSS subspectra to binarity.

%
\section{Detection efficiency} \label{sec:sims}

\begin{figure*}
 \includegraphics[width=4.8cm]{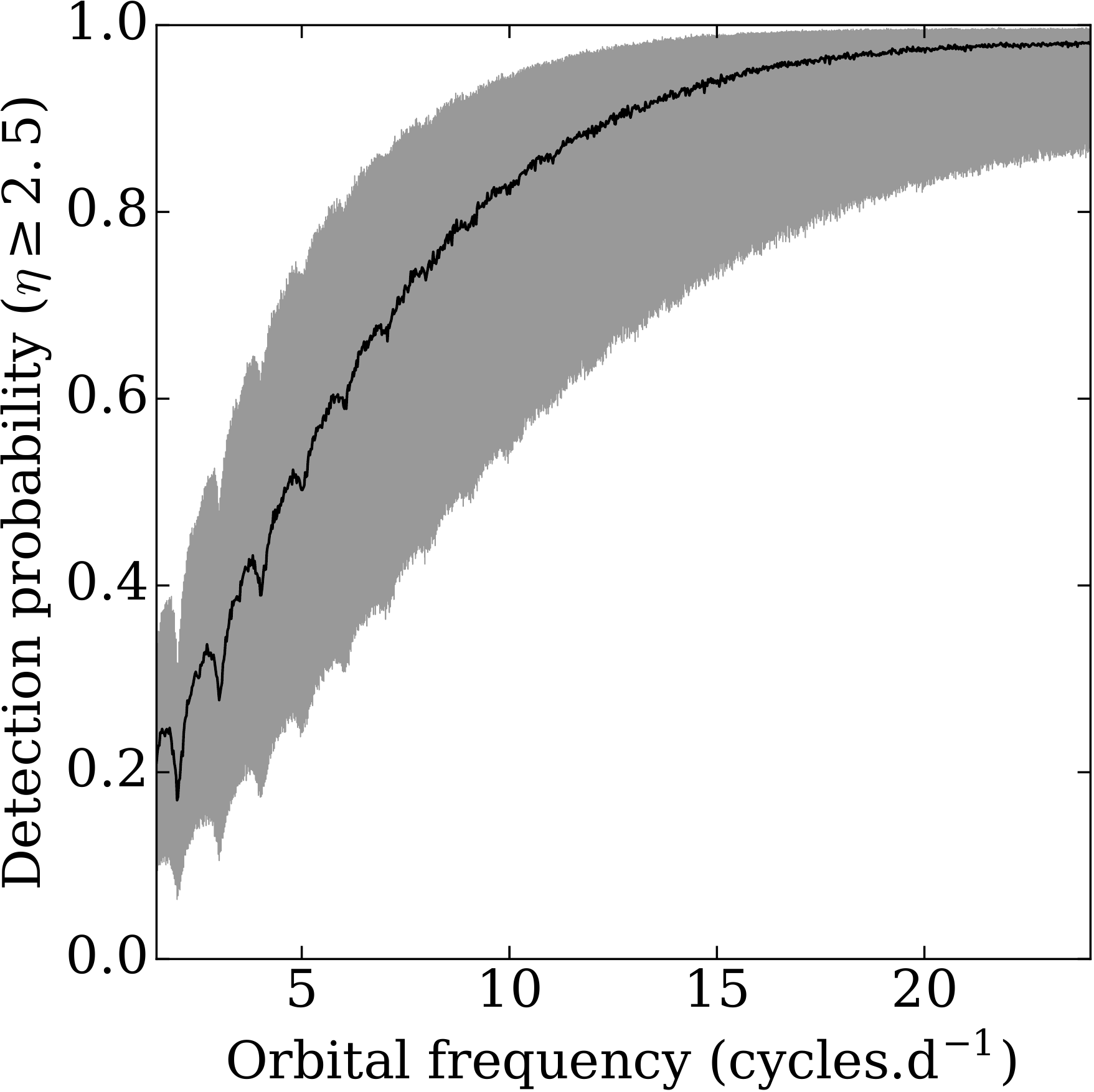}
 \hspace{2mm} 
 \includegraphics[width=6.0cm]{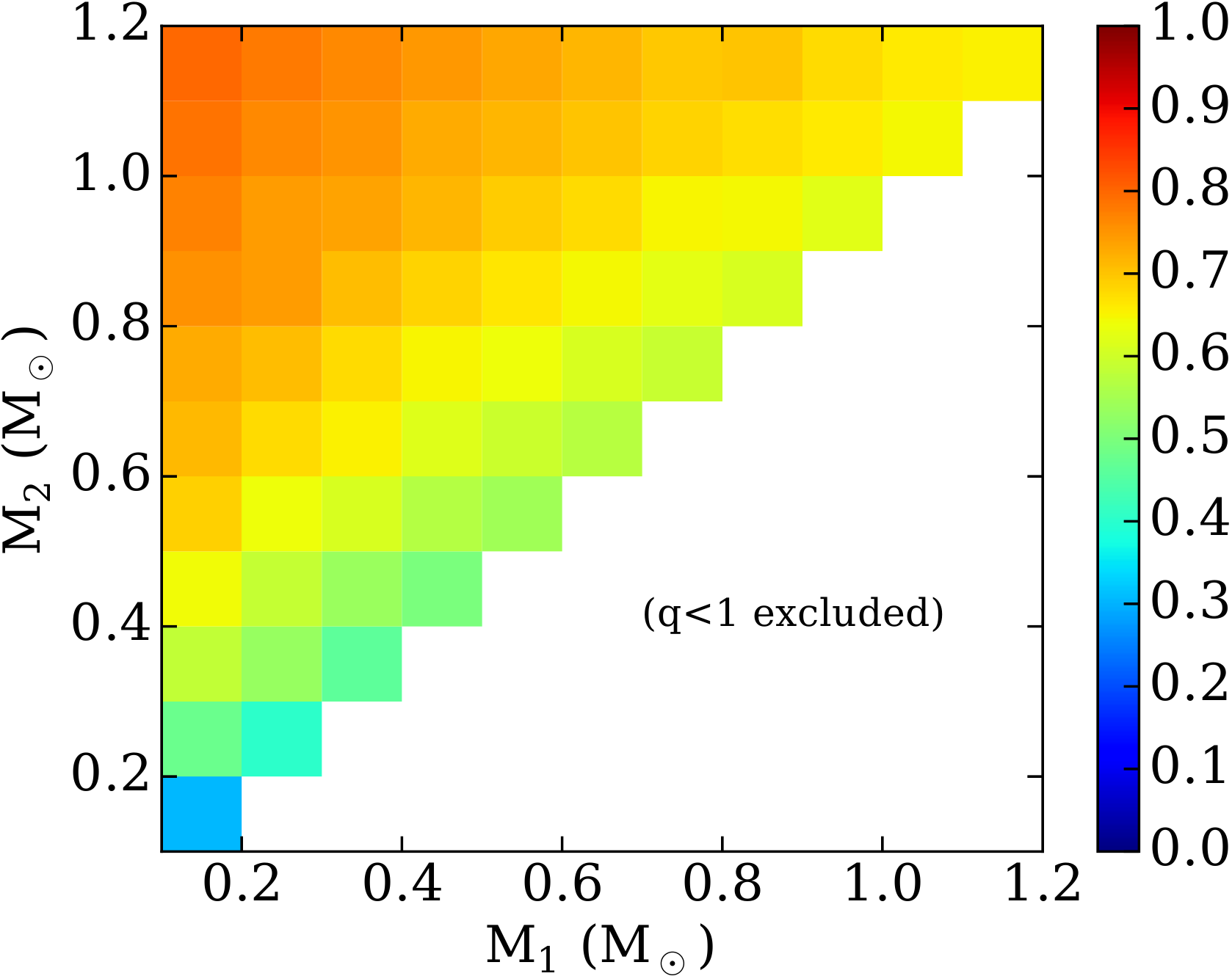}
 \hspace{2mm} 
 \includegraphics[width=6.1cm]{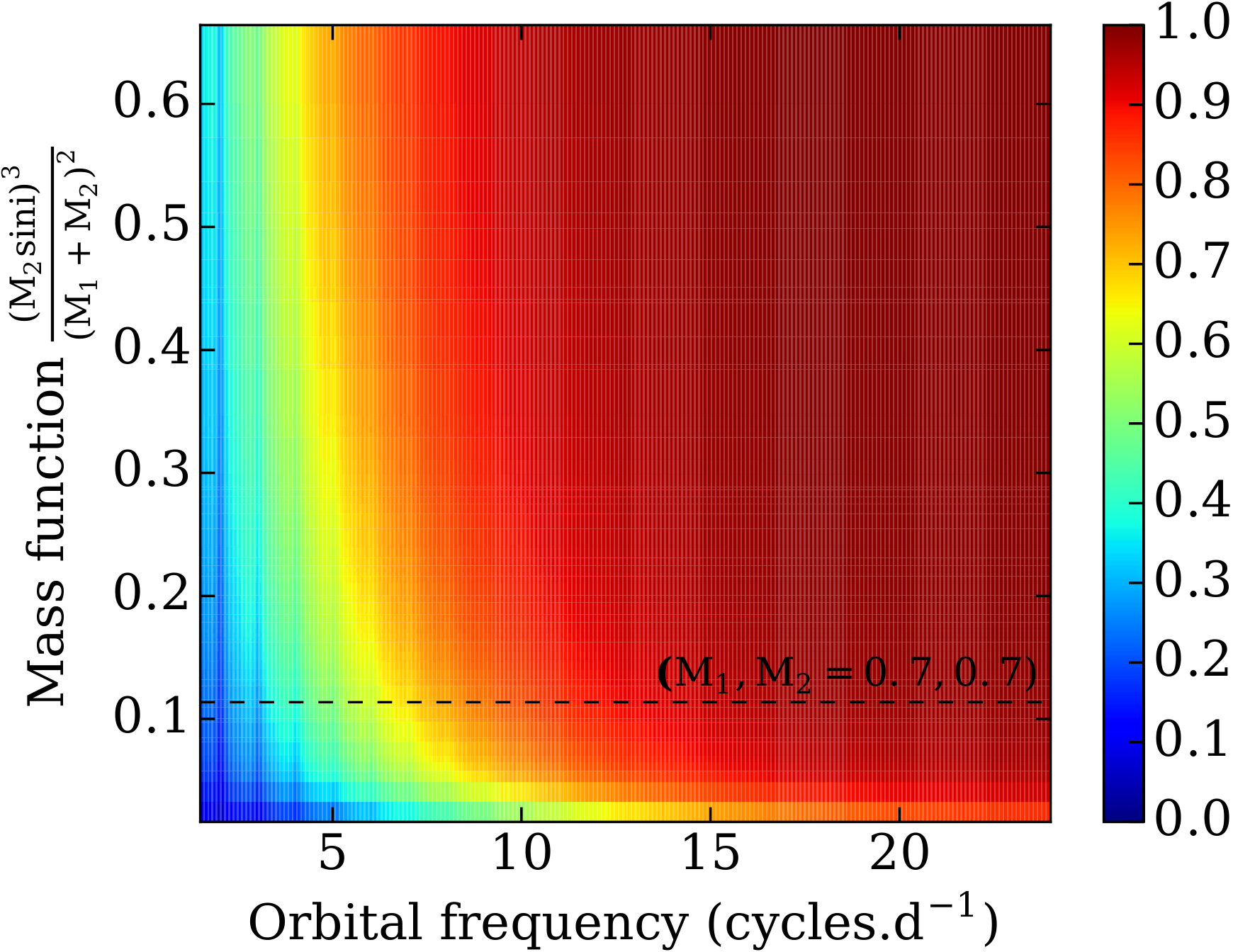}
 \caption{Detection probability as a function of the binary parameters. {\em Left:} The detection probability increases with orbital frequency. The grey area shows the range for all binary pairs in our simulation, and the $M_1, M_2 = 0.7, 0.7 M_\odot$ case is highlighted in black. {\em Middle:} Detection probability (colour scale) as a function of the masses of the component stars. The frequency was fixed at $f=6$\,cycles\,day$^{-1}$. {\em Right:} The binary mass function and orbital frequency together gives the most general overview of the detection probability. The dashed line indicates the detection probability for a $M_1, M_2 = 0.7, 0.7 M_\odot$ binary, corresponding to the black line shown in the left hand panel.}
 \label{fig:detprob}
\end{figure*}

For binaries that are potential SN\,Ia progenitors the component masses have to be large (so that the merger product exceeds or is close to the Chandrasekhar mass) and the orbital period short ($\Porb\la12$\,h; so that the merger takes place within a Hubble time). Their radial velocity amplitudes will therefore be large, and it is not unreasonable to expect that we would be able to detect such systems in the SDSS data. 

In this section, we show the results of Monte Carlo simulations to quantify the probability of detecting various types of double white dwarf binaries using this method. We started by calculating the expected orbital velocity of a white dwarf in a binary, from Kepler's third law, %
\be  v_1 = \frac{M_2}{M_1+M_2}[G(M_1+M_2)M_\odot\,2\pi f]^{\frac{1}{3}} \ee  
where $M_1$ is the bright white dwarf which dominates the optical spectrum and $M_2$ is the more massive unseen component. For a given binary pair with $M_1 < M_2$, we calculated the expected radial velocity for 1000 equally spaced orbital frequencies between $1.5\leq f\leq24$ cycles\:d$^{-1}$ (i.e. $1\leq\Porb\leq16$\,hr). We repeated the simulation for 10 values of the binary inclination $i$ for each target, randomly selected from a $\sin\,(i)$ probability distribution. We found that the inclination had only a very minor effect on our final results, so to simplify the direct comparison of the figures described below, we fixed the inclination to $i=60\degr$. We then sampled the projected velocity $K_1 = v_1 \sin i$ in the same way as the SDSS subspectra of each of the 6396 targets, adding a velocity scatter similar to the measured velocity errors as well as a random phase. 
For each case we calculated the resulting variability parameter $\eta$ and recorded the number of times it exceeded a certain threshold for the particular combination of $M_1, M_2$ and $f$. The detection probability is then the fraction of simulations for which the variability exceeded the threshold. In Figure~\ref{fig:detprob} we adopt a threshold value of $\eta=2.5$; we will return to the choice of this value in Section~\ref{sec:discussion}. 

Let us first consider the effect of orbital frequency on the detection probability (Figure~\ref{fig:detprob}, left panel). The results from all simulated mass pairs fall within the grey shaded area, and the specific case of a canonical SN\,Ia progenitor binary with $M_1, M_2 = 0.7, 0.7 M_\odot$, is highlighted in black. The dips in the probability curve at low frequencies are due to the typical observing pattern, as binaries with orbital periods of close to a day will be observed at the same phase each night, making it more difficult to identify them as variable. The detection probability increases with orbital frequency for all mass pairs considered in our simulations. 

In the middle panel of Figure~\ref{fig:detprob} we show how the detection probability varies with the masses of the component stars. We assume here that the component dominating the spectrum is the least massive of the two stars.
Mass ratios $q=M_1/M_2<1$ are excluded from the simulations. We fixed $f=6$\,cycles\:d$^{-1}$, where the detection probability in the left hand panel shows the steepest change. It shows that binaries with a more massive companion or a more extreme mass ratio have a higher probability of being detected. 

Finally in the right hand panel, we combine these results and show how the detection probability varies with  both frequency and mass, the mass dependence included in the form of the mass function, %
\be f_m = \frac{M_2^3 \sin^3 i}{(M_1+M_2)^2} = \frac{\Porb K_1^3}{2\pi G} \label{eq:massfn}. \ee %
Again we highlight the position of a $M_1, M_2 = 0.7, 0.7 M_\odot$ binary. The dashed line here corresponds to the black line shown in the left hand panel. The most obvious feature of this plot is the large range of parameters for which the detection probability is high. This again confirms that even though the sampling is sparse and the spectral resolution is much lower than SPY, SDSS can be used to uncover double degenerate binaries, particularly those with short orbital periods.


%
\section{Follow-up observations}  \label{sec:obs}

We carried out follow-up observations of targets with a range of $\eta$-values to test the reliability of the selection method. In general we prioritised the targets with highest $\eta$ that were observable at the time of our observations, but we also observed a few low $\eta$ targets which have low signal-to-noise SDSS subspectra to be able to improve their mass and radial velocity measurements. We made no selection based on the white dwarf mass, our aim was simply to confirm the variability and verify the selection method.

\subsection{Spectroscopy}  \label{sec:specdata}
Our follow-up observations involved a number of different telescopes: the 4.2\,m William Herschel Telescope (WHT; 2013 January and 2015 August) and the 2.5\,m Isaac Newton Telescope (INT; 2015 October), both at the Roque de los Muchachos Observatory on La Palma, Spain, the 6.5\,m Magellan Baade Telescope at Las Campanas Observatory, Chile (2016 May), as well as the European Southern Observatory's 8.2\,m Very Large Telescope (VLT; 2015 May and 2016 June) at Paranal Observatory, Chile. 

On the WHT we used the ISIS dual-beam spectrograph with the R600R and R600B gratings to cover the wavelength ranges $3940-5170$\,\AA\, and $5615-7140$\,\AA\, respectively. This covers \hd$\rightarrow$\ha\, at a resolution of 1.76\,\AA. The INT observations with the R632V grating on Intermediate Dispersion Spectrograph (IDS) covered \hd$\rightarrow$\hb\, ($3560-6065$\,\AA) at a resolution of 1.94\AA. For the VLT observations we selected the GRIS1200R+93 grism on the FORS2 spectrograph. This setup delivers a 1.72\AA\, resolution across the $5940-7110$\,\AA\, but it covers only the \ha\, line.  The MagE spectra from Magellan Baade have the highest resolution of our observations -- 1.6\AA\, over a wide wavelength range ($3100-10000$\,\AA). 

The overall strategy was to take two to four radial velocity snapshots separated by a few hours or nights to confirm the suspected radial velocity variability. Where possible, we took further observations of the confirmed binaries to measure their orbital periods. Unfortunately 10 out of 17 allocated nights were completely lost due to poor weather, and some of the remaining time was affected by cloud cover and/or poor seeing. Not all targets could be observed twice and some of the suspected long-period binaries need additional observations to be able to confirm the variability. The observing log shown in Table~\ref{tab:obslog} details the observations of the 36 targets which have we have obtained at least two epochs so far. These observations are ongoing; here we focus only on what the initial observations can tell us about the methodology.

\begin{table*} 
  \centering
  \caption{\label{tab:obslog} Log of follow-up observations. The temperature and $\log g$ measured from the SDSS spectra are shown for reference.}
  \begin{tabular}{cclccccrc} \hline
     Target ID   &  $g$    & UT date at     & Telescope/ & Number of & Exposure time           & $\eta$ & $\Teff$ & $\log g$ \\
 SDSS (J2000.0)  &  (mag)  & start of night & Instrument & spectra   & per spectrum (s)        & (SDSS) & (K)     &          \\ 
    \hline\hline
     J000034.07$-$010820.0 & 17.84 & 2015 Aug 11        & WHT/ISIS   & 2       & 900           &  1.07 &  13006\,(222) & 8.03(05)\\
                ''   	   &   ''  & 2015 Oct 11,12     & INT/IDS    & 7+7     & 1200          &   ''  &      ''       &    ''   \\
     J000757.81$-$050251.5 & 18.57 & 2016 Jun 18,21,22  & VLT/FORS2  & 5+9+11  & 1200          &  2.71 &  11173\,(141) & 7.90(06)\\
     J003641.60$+$242205.0 & 17.72 & 2013 Jan 3,4,5     & WHT/ISIS   & 1+1+1   & 1000--1200    &  1.95 &    7899\,(58) & 8.13(08)\\
     J022446.29$-$074820.5 & 17.72 & 2013 Jan 2,3,4     & WHT/ISIS   & 1+1+1   & 1100--1500    &  4.47 &  19216\,(160) & 7.50(02)\\
     J022919.70$+$255536.7 & 18.88 & 2013 Jan 4,5       & WHT/ISIS   & 1+1     & 1800          &  0.43 &  22262\,(462) & 7.86(06)\\
     J031524.02$+$394836.4 & 18.68 & 2013 Jan 2,3,4,5   & WHT/ISIS   & 1+1+1+1 & 1800          &  1.68 &  20024\,(370) & 7.81(06)\\
     J073616.20$+$162256.4 & 16.31 & 2015 May 7         & VLT/FORS2  & 17      & 300           &  1.64 &  20518\,(135) & 7.24(02)\\
     J075345.74$+$333527.8 & 18.59 & 2013 Jan 3,4,5     & WHT/ISIS   & 2+1+1   & 1500--1800    &  1.97 &    7339\,(89) & 7.81(15)\\
     J080004.71$+$455500.0 & 18.83 & 2013 Jan 3,4,5,6   & WHT/ISIS   & 2+1+1+1 & 1800--2000    &  1.28 &  16762\,(439) & 8.50(07)\\
     J080024.15$+$183952.0 & 18.06 & 2013 Jan 3,5,6     & WHT/ISIS   & 1+1+1   & 1100--1500    &  0.48 &  16207\,(240) & 8.09(04)\\
     J080911.20$+$352756.9 & 15.16 & 2013 Jan 3,5,6     & WHT/ISIS   & 1+1+1   & 600--1000     &  1.28 &    9091\,(19) & 8.32(02)\\
     J082001.30$+$383435.0 & 16.60 & 2013 Jan 3,5,6     & WHT/ISIS   & 1+1+1   & 900--1600     &  1.02 &    7592\,(22) & 8.01(03)\\
     J085858.49$+$301231.3 & 17.69 & 2013 Jan 3,4,5,6   & WHT/ISIS   & 1+1+1+1 & 1000--1800    &  0.37 &  31697\,(267) & 8.02(06)\\
     J090014.38$+$331140.0 & 18.54 & 2013 Jan 4,5,6     & WHT/ISIS   & 1+1+1   & 1300--1500    &  0.90 &    9787\,(94) & 8.47(09)\\
     J093009.37$+$420754.1 & 17.47 & 2013 Jan 3,5,6     & WHT/ISIS   & 1+1+1   & 900--1800     &  0.62 &  16334\,(164) & 7.88(03)\\
     J094652.19$+$631841.7 & 18.81 & 2013 Jan 4,5,6     & WHT/ISIS   & 1+1+1   & 1500--1800    &  4.30 &  10939\,(201) & 8.03(10)\\
     J100628.33$+$624815.0 & 18.49 & 2013 Jan 3,4,5,6   & WHT/ISIS   & 1+1+1+1 & 1200--1600    &  0.16 & 44289\,(1868) & 7.86(19)\\
     J101035.46$+$262946.9 & 18.79 & 2013 Jan 4,6       & WHT/ISIS   & 1+1     & 1100, 1500    &  0.38 &  32048\,(501) & 7.07(12)\\
     J112721.28$-$020837.4 & 16.21 & 2015 May 5         & VLT/FORS2  & 8       & 180           &  4.15 &   26720\,(99) & 7.86(02)\\
     J113709.83$+$003542.7 & 18.46 & 2013 Jan 4,5,6     & WHT/ISIS   & 1+1+1   & 1200--1400    &  0.79 &  17886\,(396) & 7.63(07)\\
     J122450.26$+$003617.3 & 18.57 & 2013 Jan 4,6       & WHT/ISIS   & 1+1     & 1800, 1400    &  0.43 &  30001\,(716) & 7.42(15)\\
     J123549.86$+$154319.0 & 17.34 & 2013 Jan 3,4       & WHT/ISIS   & 4+1     & 800--1000     &  26.50 & 22096\,(235) & 7.21(03)\\
     J125714.33$+$184516.9 & 17.23 & 2015 May 5         & VLT/FORS2  & 7       & 600           &  0.22 &  13730\,(230) & 8.10(03)\\
	 J133336.04$+$091704.9 & 17.93 & 2015 May 5         & VLT/FORS2  & 4       & 600           &  2.23 &  14026\,(349) & 8.30(05)\\
     J141516.08$-$010912.1 & 18.27 & 2016 Jun 18,21,22  & VLT/FORS2  & 11+11+17 & 900          &  5.27 &    6106\,(60) & 8.12(13)\\
     J144108.42$+$011019.9 & 16.88 & 2015 May 5         & VLT/FORS2  & 4       & 300           &  2.28 &  32636\,(162) & 7.32(04)\\
     J144510.28$+$141344.9 & 16.12 & 2015 May 5         & VLT/FORS2  & 2       & 150           &  12.40 & 50481\,(593) & 6.90(05)\\
     J162314.70$+$142829.0 & 18.88 & 2016 Jun 18,22     & VLT/FORS2  & 1+1     & 1500          &  4.82 &   9674\,(126) & 7.84(14)\\      
     J162858.17$+$141508.5 & 17.21 & 2015 May 5         & VLT/FORS2  & 2       & 600           &  1.18 &   10335\,(51) & 8.00(04)\\
     J163133.15$+$153329.3 & 18.24 & 2016 Jun 18,21     & VLT/FORS2  & 6+4     & 900           &  2.78 &    9265\,(84) & 8.00(10)\\   
     J163826.31$+$350011.9 & 14.64 & 2015 Oct 10,11,12  & INT/IDS    & 1+2+1   & 1800--2700    &  5.33 &  37787\,(180) & 7.90(02)\\
     J171125.53$+$272405.1 & 17.19 & 2015 May 5         & VLT/FORS2  & 1       & 600           &  2.59 &  54091\,(1351)& 7.94(09)\\
               ''          &   ''  & 2016 May 21        & VLT/FORS2  & 2       & 1440          &   ''  &      ''       &    ''   \\
     J180943.69$+$235828.8 & 17.51 & 2015 May 5         & VLT/FORS2  & 3       & 600           &  0.95 &  19782\,(200) & 7.97(03)\\
     J204544.80$-$001614.5 & 18.19 & 2015 May 5         & VLT/FORS2  & 2       & 600           &  1.57 &   6258\,(168) & 8.98(34)\\
               ''          &   ''  & 2016 May 21,22     & VLT/FORS2  & 13+12   & 720           &   ''  &      ''       &    ''   \\
     J231951.69$+$010908.3 & 18.46 & 2013 Jan 2,3,4     & WHT/ISIS   & 1+1+1   & 1800          &  1.20 &   7167\,(131) & 8.72(20)\\
     J234902.80$+$355301.0 & 18.42 & 2013 Jan 2,3,4,5   & WHT/ISIS   & 1+1+3+8 &  900--1800    & 19.94 &  25495\,(444) & 7.25(06)\\
     J235552.69$+$372143.5 & 18.86 & 2013 Jan 2,3,4     & WHT/ISIS   & 1+1+1   & 1600--1800    &  0.58 &  16870\,(291) & 7.98(06)\\
\hline 
\end{tabular}
\vfill
\end{table*}


\subsection{Results}  \label{sec:results}

\begin{table*} 
  \centering
  \caption{\label{tab:rvs} Radial velocity measurements from follow-up observations (Extract only. Full table available online.)}
  \begin{tabular}{llrrrrrrr} \hline
     Target ID   &  BTDB$-2450000$    & Radial velocity     & Velocity err & $\eta$      & $\eta$ & $\eta$ \\
 SDSS (J2000.0)  &  (mid-exp)         & (\kms)              & (\kms)       & (follow-up) & (SDSS) & (both) \\ 
    \hline\hline
J123549.86$+$154319.0
    & 3442.401166  &     16.0  &   23.4  & & \\
    & 3442.412613  &    -88.4  &   24.2  & & \\
    & 3442.423997  &    164.9  &   23.9  & & \\
    & 6013.304119  &   -102.3  &   20.8  & & \\
    & 6013.315508  &     53.4  &   20.4  & & \\
    & 6013.326874  &    168.9  &   20.1  & & \\[4pt]
    & 6295.242296  &   -143.0  &   21.7  & & \\
    & 6295.252803  &    -51.8  &   23.4  & & \\
    & 6295.264123  &    129.4  &   24.0  & & \\
    & 6295.273473  &    -73.7  &   22.6  & & \\
    & 6296.273024  &   -128.9  &   22.9  & 17.12 & 26.50 & 54.00\\[10pt]
\hline
\end{tabular}\\
\end{table*}

\begin{figure}
 \rotatebox{0}{\includegraphics[width=1.05\columnwidth]{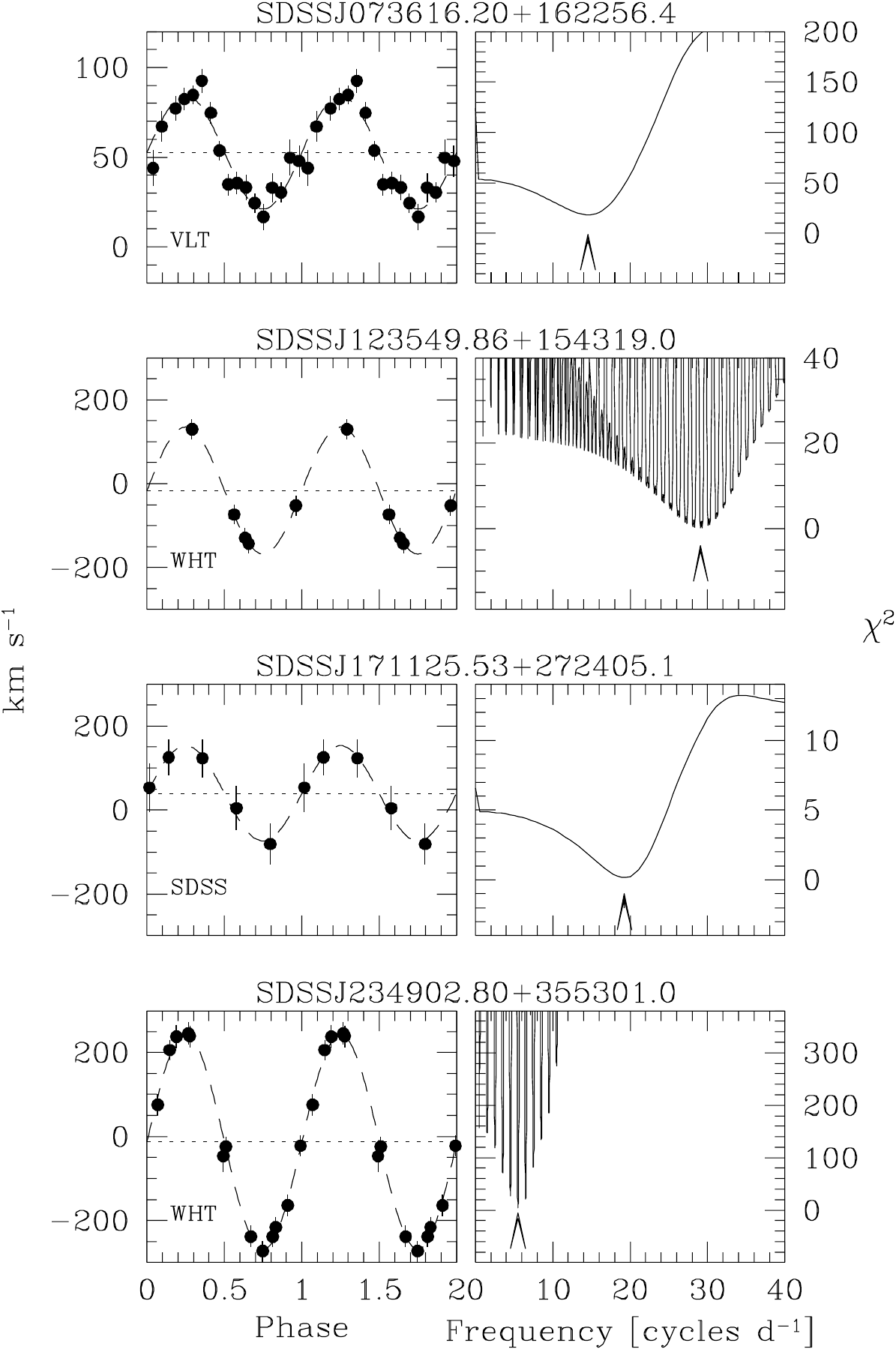}}
 \caption{Radial velocity measurements folded on the period determined from the lowest $\chi^2$ fit of a circular orbit. The adopted orbital frequency is indicated by an arrow and the parameters of the fit are shown in Table~\ref{tab:newbinaries}. The radial velocities of SDSS\,J171125.53$+$272405.1 are from the SDSS subspectra only; the other targets are labelled with the name of the telescope where the observations were taken.}
 \label{fig:rvperiods}
\end{figure}

We measured the radial velocities in exactly the same way as for the SDSS spectra, using a simultaneous multiple Gaussian fit to all the Balmer lines covered by the setup. The radial velocities are shown in Table~\ref{tab:rvs}, including the $\eta$ measured from the SDSS as well as the follow-up spectra. The table is available in full in the online material.  We select $\eta=4$ in the follow-up data as the threshold for confirming a target as a double degenerate binary. For this value, the number of false alarms in our sample, simply due to Gaussian statistics, is $<1$ (See Section~\ref{sec:discussion}). In general our follow-up observations confirm that the targets with large values of $\eta$ are radial  velocity-variable. As mentioned previously, there are several targets for which the follow-up observations do not cover a long enough baseline, so our observations are insufficient to make any conclusive statement of the variability of these targets. Seven of our targets exceed $\eta\geq4$; we highlight these with bold print in the online table. For four of these binaries we obtained enough measurements to measure their orbital periods (Figure~\ref{fig:rvperiods}). 
We assume that the orbits are circular and fit the data with a series of sine waves, 
\be V(t) = \gamma + K_1 \sin[2\pi(t-\mbox{HJD}_0)f] \label{eq:orbit} \ee %
equally spaced in frequency. We adopt the frequency resulting in the lowest $\chi^2$ of the fit as the orbital frequency of the binary. The large uncertainty on the periods of SDSS\,J073616.20$+$162256.4 and SDSS\,J171125.53$+$272405.1 result from the fact that the observations were taken on a single night. For SDSS\,J123549.86$+$154319.0 and SDSS\,J234902.80$+$355301.0 the $\chi^2$ of the second-best alias are factors of 6 and 5 larger than the minimum, respectively, so these are likely to be the correct periods of these binaries. The one-day alias periods are however still possible. The parameters of the best fit circular orbit for each of the four binaries are shown in Table~\ref{tab:newbinaries}. The orbital period of SDSS\,J123549.86$+$154319.0 turned out to be very short (59 min), so the 900\,s exposure time we used in the follow-up spectroscopy span a large fraction of the orbit. This phase smearing reduces the measured velocity amplitude by a factor %
$(\sin x)/x$ compared to the true amplitude, where $x=\pi t_{\mbox{\tiny exp}}/\Porb$ and $t_{\mbox{\tiny exp}}$ is the exposure length.  
We include both the measured ($K_1$) and corrected ($K_c$) velocity amplitudes in Table~\ref{tab:newbinaries} and derive the other parameters in that table using the corrected amplitude. In terms of finding massive white dwarf binaries, SDSS\,J234902.80+355301.0 is the most interesting of the targets. It has a total mass $M>1.115\Msun$ and a gravitational wave radiation merger time scale of $<1.9\times10^9$~years, well within a Hubble time. With only a lower limit on $M_1$ it is not yet clear whether it is massive enough to be a SN\,Ia progenitor in the classic sense, but the lower limit on the total mass already falls in the range of interest for double-detonation SN\,Ia models.  
Given the low mass functions derived for the other three binaries, we checked again for any signs that the companion might be a late type dwarf star rather than a white dwarf. First we checked whether their spectra show any infrared excess compared to the white dwarf spectrum only. Of the four binaries in Table~\ref{tab:newbinaries}, only SDSS\,J123549.86$+$154319.0 falls within the {\em UKIDSS\,}\footnote{United Kingdom Infrared Telescope (UKIRT) Infrared Deep Sky Survey} footprint. It is detected in the $Y$, $J$ and $H$ bands only, at fluxes that follow the spectrum of a 22\,100~K white dwarf exactly. The non-detection in the $K$ band is also consistent with this white dwarf spectrum. We also checked the {\em WISE\,}\footnote{Wide-field Infrared Survey Explorer} All Sky Catalogue, but none of the four binaries is detected at the {\em WISE} wavelengths. Finally, we also downloaded the Catalina Sky Survey\footnote{\citealt{drake09crts}. Public data release available from http://nesssi.cacr.caltech.edu/DataRelease/} light curves of these binaries to check for optical variability that could result e.g. from ellipsoidal modulation or irradiation of a brown dwarf companion by a hot white dwarf. We find no significant periodic variability in any of the light curves. Without near-infrared data there remains the possibility of a faint late-type companion that is below the detection limit at {\em WISE} wavelengths, but the data available at the moment is consistent with the companions being white dwarfs.

The subspectra of SDSS\,J144510.28+141344.9 ($\Twd=56\,000$~K; $\log g=6.82$) display significant, but likely long period radial velocity variations. We were only able to obtain two consecutive exposures of this target, so we could not confirm the variability, but we nevertheless include it here. The two VLT spectra display broad and variable profiles of the \ha\, absorption core, along with what appears to be a weak \ha\, emission line. This suggests that the white dwarf might be accreting, but as above, the target has no infrared excess emission that might indicate the presence of a low mass companion and no periodic variability in its Catalina Sky Survey light curve. Instead we suggest that this might be a double-lined white dwarf binary, showing absorption from both the component stars. We encourage further high resolution observations of this target to confirm this and to measure its mass ratio.

One of our follow-up targets, SDSSJ112721.28$-$020837.4 = WD1124$-$018, was observed by SPY as well. The SPY spectra display a velocity change of $\Delta v_{\mbox{\small max}} = 101.9$\,\kms\, over $\Delta t =1.9$~days, identifying the target as a double degenerate binary \citep{maoz17}. Our VLT observations have $\eta=3.88$, placing it just outside the threshold where we would assume it to be a double degenerate without further follow-up. The VLT observations have $\Delta v_{\mbox{\small max}} = 32.7$\,\kms\, between spectra taken $\Delta t =138$~minutes apart, which would be sufficient to include it among the list of very likely double degenerate binaries by the $\Delta v_{\mbox{\small max}}>15$\,\kms\, criterion that  \citet{maoz17} use.

\begin{table*} 
  \centering
  \caption{\label{tab:newbinaries} Orbital parameters of the new binaries. $M_1$ and $\Teff$ are derived from the atmosphere model fit. $M_2$ is the minimum mass derived from the mass function.}
\begin{tabular}{lccccccccl} \hline
     Target ID              &  $\Porb$   &  $T_0$        &   $K_1$ & $K_c$  &  $\gamma$  & $M_1$     & $f_m$     & $M_2$     & $\Teff$ \\
      SDSS (J2000.0)        &  (d)       &(HJD$-2450000$)&  (\kms) & (\kms) &  (\kms)    & ($\Msun$) & ($\Msun$) & ($\Msun$) & (K)     \\
    \hline\hline
     J073616.20$+$162256.4  & 0.069(4)   & 7150.533(1)   & 32(2)   &  32(2)   &  53(2)     & 0.360(7)  & 0.0002 & $>0.033$ & 20\,518 \\
     J123549.86$+$154319.0  & 0.03438(5) & 6296.2680(9)  & 151(18) &  176(21) &  $-16(12)$ & 0.363(11) & 0.0196 & $>0.179$ & 22\,096 \\
     J171125.53$+$272405.1  & 0.052(7)   & 5715.374(4)   & 113(32) &  121(34) &  40(22)    & 0.672(44) & 0.0095 & $>0.192$ & 54\,091 \\
     J234902.80$+$355301.0  & 0.1813(4)  & 6296.925(3)   & 255(10) &  257(10) &  $-13(8)$  & 0.386(14) & 0.3207 & $>0.742$ & 25\,495 \\
\hline\\ 
\end{tabular}\\
\end{table*}


%
\section{Discussion} \label{sec:discussion}

\subsection{Effective sample size}

\begin{figure}
 \centering
 \rotatebox{0}{\includegraphics[width=0.9\columnwidth]{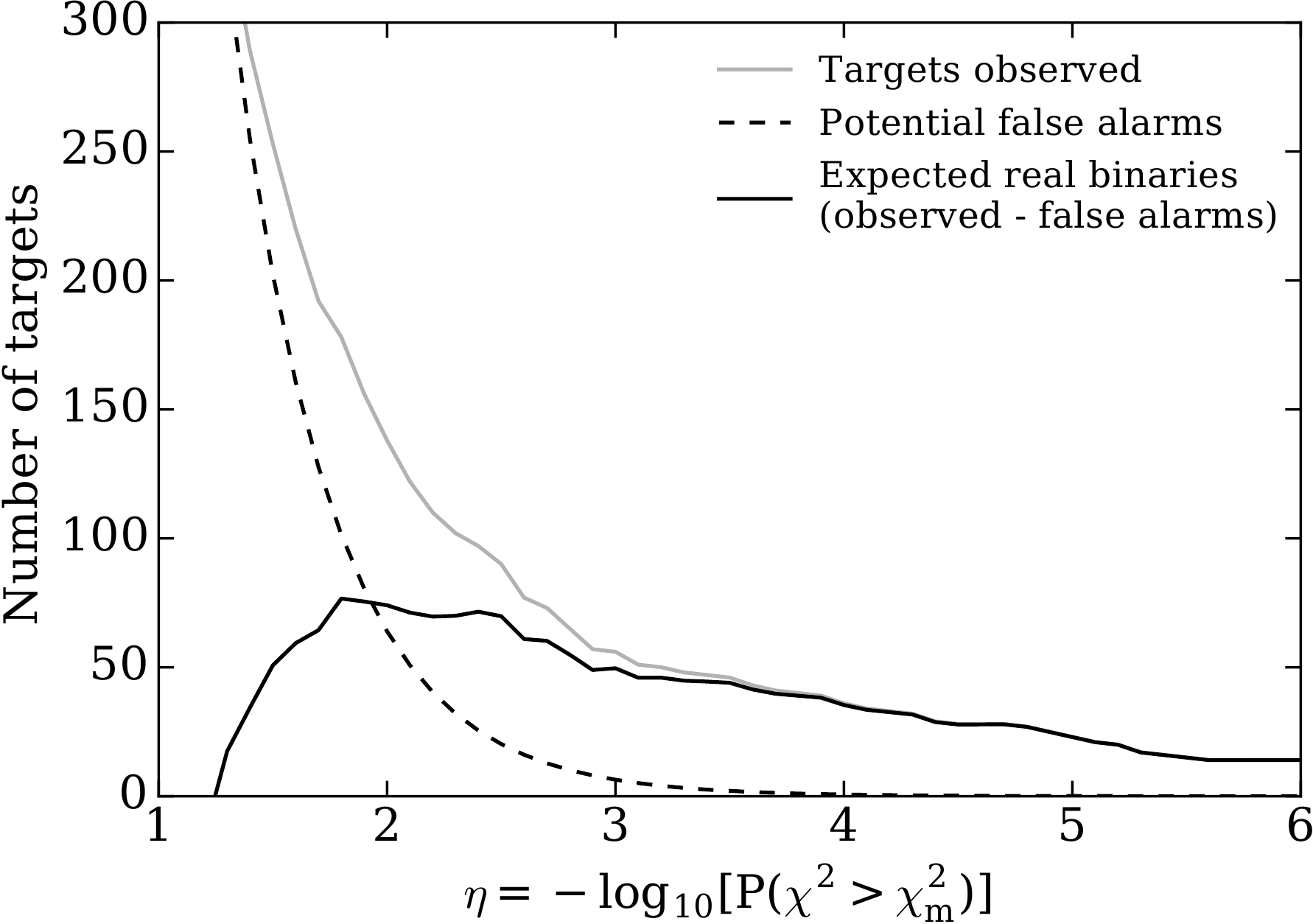}}
 \caption{Number of targets to be observed as a function of the variability threshold $\eta$ (grey), compared to the number of false alarms we can expect based on Gaussian noise (dashed line). The black solid line is the difference between the two, showing that for $\eta>4$ the target is a double degenerate to high confidence. $\eta=2.5$ gives good balance between the effective survey size and the number of false positives. }
 \label{fig:whicheta}
\end{figure}

Potential SN\,Ia progenitors comprise only a small fraction of the total population of double degenerate binaries. As discussed in the introduction, the Galactic SN\,Ia rate requires that at least one in every $\sim400$ white dwarfs we test for radial velocity variability should be a SN\,Ia progenitor, if indeed merging double degenerates are responsible for most SN\,Ia explosions. So, to test whether the classic double degenerate model is correct, we need to sample at least several times this number of white dwarfs. SPY surveyed 1000 targets, but many had to be excluded because they turned out to be subdwarf stars rather than white dwarfs, because the white dwarf was found to be magnetic (the Zeeman splitting of the lines complicates the radial velocity measurement), or because only one spectrum could be obtained during the period of observations. In total 615 DA white dwarfs were surveyed for radial velocity variability --- this is the {\em survey size} of the SPY survey  \citep{koester09}.

The advantage of pre-selecting or ranking the targets using the SDSS velocities, is that non-variable (single) white dwarfs can be reliably excluded from the follow-up sample, leading to a much reduced cost in observing time. A survey based on the variability ranked sample only has to spend observing time on the targets with $\eta$ above a chosen threshold, and this observed sample will contain a much higher binary fraction than the sample as a whole. The detection probability allows us to account for the real binaries that are excluded along with the single white dwarfs, due to the sparse sampling pattern and spectral resolution of the SDSS data. 

In a survey where no prior information about the variability is available, all the candidates have to be followed up, so the survey size is the same as the number of observed targets, at best. In a ranked survey, we define the {\em effective survey size} as the fraction of binaries we can expect to detect from the whole sample, i.e. the full number of targets for which we have radial velocity information (6396 in the case of SDSS) multiplied by the detection probability of a given binary in that survey. The detection probability also depends on the threshold $\eta$ down to which we are willing to carry out follow-up observations. Setting a high threshold means that only those targets for which we detected variability with high confidence in the SDSS subspectra will be followed up, but observing so few targets from the full sample reduces the overall detection probability of a given binary. Conversely, setting the threshold too low will result in a higher detection probability, but it means that most of the sample will have to be followed up. Also, from the definition of $\eta$ (equation~\ref{eq:eta}), the lower $\eta$ the larger the number of false positives we can expect, so the best choice of $\eta$ is a trade-off between the number of targets to be observed, the detection efficiency and the expected number of false positives. For the SDSS sample, we find that $\eta=2.5$ strikes a good balance (Figure~\ref{fig:whicheta}). For $\eta=3.0,\: 2.5$ and $2.0$ we will have to observe 56, 90 and 138 targets respectively, of which we may expect 6, 20 and 64 to be false positives, according to Gaussian statistics.

\begin{figure}
 \centerline{{\includegraphics[width=0.95\columnwidth]{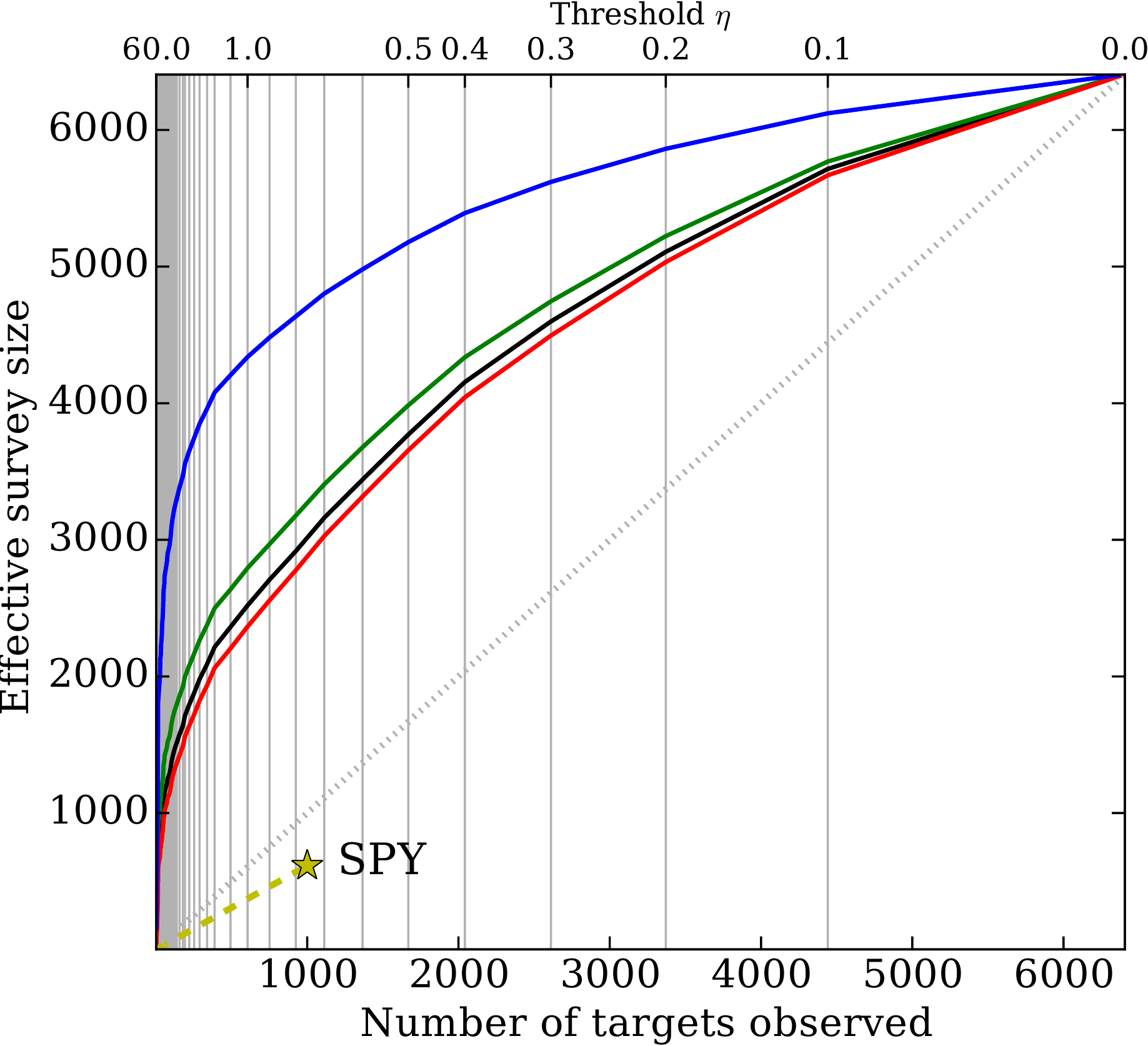}}}
 \vspace{2mm}
 \centerline{{\includegraphics[width=0.95\columnwidth]{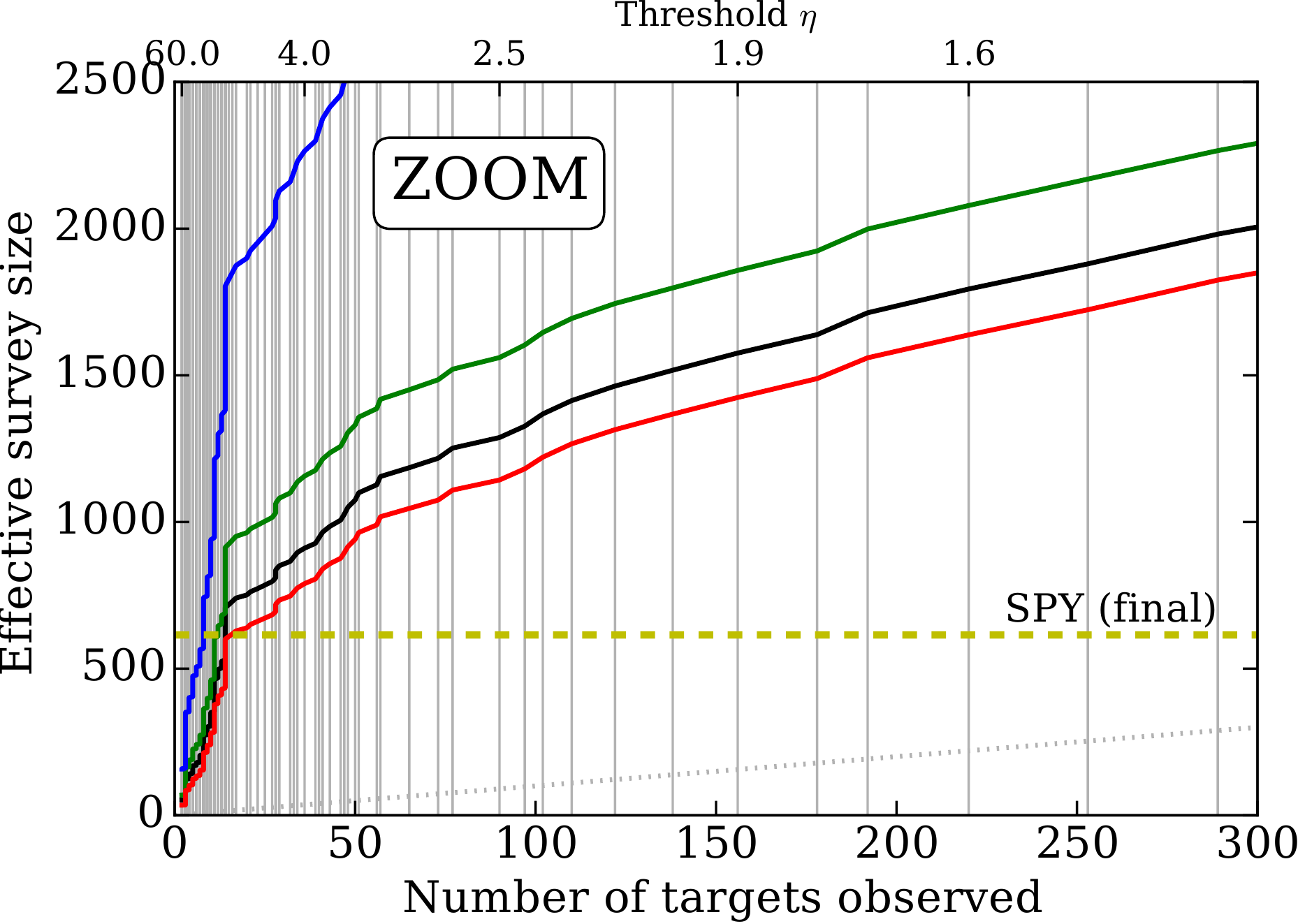}}}
 \caption{Comparison between the effective survey size of a ranked dataset (coloured solid lines) with a blind survey (grey dotted line) and SPY (yellow dashed line). The effective sample size depends on the detection probability, so varies with the parameters of the simulated binary. The red, black and green lines are the long-period ``pessimistic'' cases, representing a $M_1,M_2 = 0.7,0.7$ (black), $M_1,M_2 = 0.6,0.6$ (green) and $M_1,M_2 = 0.6,0.8$ (red) binary just merging within a Hubble time ($\Porb\sim11$\,hr). The detection probability of a shorter period binary is higher, so for comparison, a $M_1,M_2 = 0.6,0.8$ binary with $\Porb=4$\,hr is shown in blue. The top panel shows the full sample of 6396 white dwarfs. The bottom panel shows an expanded version of the $x$-axis to illustrate rapid increase in the effective survey size for a small number of targets observed.}
 \label{fig:Neff}
\end{figure}

To illustrate how these parameters affect the effective survey size and how ranking increases the efficiency of the survey, we show a few specific cases in Figure~\ref{fig:Neff}. The horizontal axis represents the number of targets from the ranked SDSS sample for which we have to obtain follow-up observations, i.e. the number of targets which have $\eta$ above the threshold indicated by the grey vertical lines. For each choice of $M_1, M_2, \Porb$ and $\eta$, we can calculate the detection probability and hence the effective survey size. This is indicated on the vertical axis. A 1:1 correspondence is plotted as a dotted line, representing the case of a survey where all targets are known white dwarfs, and targets are followed up in random order, as no prior information of their radial velocity variability is available. This is in principle the technique that SPY used, but because of the interlopers in the survey, its efficiency was somewhat below this level, indicated by a yellow dashed line and a star symbol for the final survey statistics . We compare these values with the most pessimistic scenario for a SN\,Ia progenitor: a combined mass of $M_{\mathrm{Ch}}$ and an orbital period long enough that the binary will only just merge in a Hubble time. 
We show three examples in Figure~\ref{fig:Neff}. The case $M_1, M_2 = 0.7,0.7\Msun$ is plotted in black, $M_1, M_2 = 0.6,0.8\Msun$ is plotted in red and a sub-Chandrasekhar binary with $M_1, M_2 = 0.6,0.6\Msun$ is shown in green. For these types of binaries our detection probability with SDSS is low. Yet, even so, the effective sample size far exceed that of SPY. Consider the `worst case' shown (red line; $M_1, M_2 = 0.6,0.8\Msun$). Choosing $\eta=2.5$ as the threshold for follow-up observations, means that we will have to observe 90 targets. The detection probability in this case is very low, only 18 per cent, which gives an effective survey size of 1143. Despite the low detection probability, this is almost double the the survey size of SPY (a factor of 1.8), for observing less than 15 per cent as many targets. 
For comparison, we also show the effective survey size for a short period binary, to which SDSS has a much higher sensitivity. The blue line in Figure~\ref{fig:Neff} represents the same binary with $M_1, M_2 = 0.6,0.8\Msun$, but with an orbital period of four hours. The survey size in this case is a factor of 4.8 greater than SPY for the same $\eta$.

\subsection{Comparison with observations}
Our follow-up observations are not yet sufficient to make any detailed comparisons with the simulations, but the initial results are promising. Of the top 20 targets in our sample, as arranged by $\eta$, six are previously confirmed binaries and a further three have been confirmed by our observations. 10 of the remaining top-20 are still awaiting observations, or have only been observed once so far in our programme. Considering only the targets that have been observed, we find only one with $\eta>4$ which does not have significant variability in the follow-up observations (Table~\ref{tab:rvs}), so the method is efficient at selecting binaries. 

These initial observations also illustrate the importance of good signal-to-noise and the radial velocity uncertainty. For a binary with a high mass function an error of $\sim$20\,\kms\, may be tolerated, but in general we find that we need an error of $\lesssim10$\,\kms\, to rule out radial variability in the follow-up observations. With poor weather conditions this is not always easy to achieve. As a result of large velocity errors, some of our high $\eta$ binary candidates could not be confirmed as variable, but binarity is not ruled out by our observations either. Similarly, we included a few low $\eta$ targets in our follow-up sample, of which some were found to be variable, e.g. SDSS\,J100628.33+624815.0 and SDSS\,J122450.26+003617.3 (Table~\ref{tab:rvs}). In both cases the low $\eta$ resulted from large errors on the SDSS velocities (due to the broad, shallow lines of these white dwarfs and the low signal-to-noise of their subspectra). As discussed in Section~\ref{sec:sims}, a target with a long orbital period can also have a low $\eta$ in SDSS if all the subspectra were taken on the same night. Further epochs are needed to measure the orbital periods of these binaries. With a small number of measurements contributing to $\eta$, a single observation could have a big effect, so this should not be viewed as an absolute measure of the variability of the target. With further observations in the future, we will link these observations more closely to the simulations by calculating a follow-up `efficiency factor'. This will help us to account for effects such as poor weather or technical errors on the simulated survey efficiency. Nevertheless, even with a slightly reduced efficiency, Figure~\ref{fig:Neff} shows that only a small number of observations are necessary before the ranked survey exceeds the efficiency of a blind or random survey.

\subsection{Overlap with SPY}
The SPY and SDSS samples are highly complementary (see \citealt{maoz17} for a discussion of the binary separation distributions that the two surveys are sensitive to) and largely independent. There are only 52 white dwarfs common to both surveys, 49 of which are likely single stars. We measure $\eta<0.7$ for all these stars from the SDSS subspectra; in fact, most have $\eta<0.1$, so well within the noise for all mass ranges (Figure~\ref{fig:psplitmass}). Of the three that were identified as binaries by SPY, we identified one as a likely binary as well (WD1124$-$018; Section~\ref{sec:obs}) but the remaining two (HS1102$+$0934 and HS1204$+$0159) could not be identified as variable from their SDSS subspectra, likely due to long orbital periods. For both targets, the SDSS observations were taken within 45 minutes, while the SPY epochs are separated by six and one nights for the two binaries respectively.

\subsection{Future spectroscopic surveys}
The known population of double white dwarfs is heavily biased towards low mass binaries. Given the importance of ELM binaries for models of common envelope evolution \citep[e.g.][]{nelemans05} and the low frequency gravitational wave background \cite[e.g.][]{hermes12}, much of the observational effort for double white dwarfs is currently directed towards low mass systems. Only a few systems are known at the massive end of the distribution where we expect to find the SN\,Ia progenitors. The small size and biases in the observational input is currently one of the biggest factors limiting the accuracy of population synthesis models \citep[e.g.][]{toonen12}. Detecting and characterising massive white dwarf binaries, even if they are not super-Chandrasekhar, is very important to help eradicate this bias and provide reliable observational input to models. 

Future large-scale spectroscopic surveys such as WEAVE, DESI and 4MOST are expected to observe $\gtrsim10^5$ new white dwarfs identified by Gaia \citep{jordan07,carrasco14}. WEAVE \citep[the William Herschel Telescope Enhanced Area Velocity Explorer;][]{weave} is a new multi-object spectrograph for the 4.2-m William Herschel Telescope at the Observatorio del Roque de los Muchachos, on the island of La Palma in the Canary Islands. It will provide wide-field spectroscopic follow-up for various ground- and space-based surveys in the form of key science projects\footnote{See http://www.ing.iac.es/weave/science.html}. $1-2\times10^5$ white dwarfs will be observed at medium resolution ($R\sim5000$, velocity resolution $\sim5$\,\kms) as part of the `Galactic Place Stellar Circumstellar and Interstellar Astrophysics' (SCIP) project, in order to determine the ages of various stellar populations and study the star formation history of the Galaxy. White dwarfs will also act as flux calibrators, so will be included in other science programmes as well. The current survey strategy allows for multiple visits to the survey fields, so the WEAVE data are ideal for applying our ranking method. WEAVE is expected to see first light early in 2018 with science operations to start in mid-2018.

Similarly, while the DESI (Dark Energy Spectroscopic Instrument)\footnote{http://desi.lbl.gov/} observations will be directed towards emission line galaxies and quasars, white dwarfs will be included in each pointing to allow for flux calibration. DESI observations are scheduled to start in 2019. In the longer term, 4MOST (the 4-metre Multi-Object Spectroscopic Telescope), will run a southern sky spectroscopic survey using the VISTA telescope in Chile  \citep{4most}. The survey has Galactic Archaeology as one of the key science themes, and will observe $10^5$ white dwarfs as part of this programme. 

These data will give a much better estimate of the white dwarf binary fraction as well as tight constraints on SN\,Ia progenitor models, but it will be impossible to follow-up all targets to confirm the variability and measure the binary parameters. An efficient, statistically well-understood strategy will be essential to identifying the targets worthy of further follow up. Our aim is to test and refine the methodology using the SDSS sample, to be able to provide reliable input to future large surveys.


%
\section{Summary} \label{sec:summary}

Despite the relatively low resolution and short timescale of the observations, the SDSS spectra are a rich source of useful radial velocity information. Our Monte Carlo simulations show that by using these radial velocities to pre-select variable candidates for follow-up observations, we can efficiently survey a large number of white dwarfs in a small amount of observing time. Monte Carlo simulations show that, compared to the previous largest white dwarf radial velocity survey, SPY, this method allows us to double the survey size for less than 15 per cent of the required observations. Our follow-up observations are still ongoing, but initial observations already validate the method. An efficient survey strategy for double degenerates will be essential for the upcoming large spectroscopic surveys, which are expected to yield several hundred thousand new white dwarfs.


%
\section*{Acknowledgements}
Based on observations made with ESO Telescopes at the La Silla Paranal Observatories under programmes 095.D-0837(A), 095.D-0965(A) and 097.D-0663(A), as well as observations with the William Herschel and Isaac Newton Telescopes at the Roque de los Muchachos Observatory, La Palma, Spain. 
Financial support for this work was provided by the Science and Technology Research Council (EB, DS and TRM; grant ST/L00073), the European Research Council (NPGF, PET and BTG; grant numbers 677706-WD3D and 320964-WDTracer), the NASA  Hubble Fellowship programme awarded by the Space Telescope Science Institute (JJH; grant \#HST-HF2-51357.001-A), the Joint Committee ESO--Government of Chile (MB; grant 2014), the Ag\`encia de Gesti\'o d'Ajuts Universitaris i de Recerca (ARM) and MINECO grant AYA2014-59084-P (ARM).
Funding for Sloan Digital Sky Survey has been provided by the Alfred P. Sloan Foundation, the Participating Institutions, the National Science Foundation, and the U.S. Department of Energy Office of Science. 

\bibliographystyle{mnras}
\bibliography{libraryDWD}

\bsp

\label{lastpage}

\end{document}